\documentclass[prb,twocolumn,showpacs,preprintnumbers,amsmath,amssymb,longbibliography]{revtex4-1}
\usepackage{graphicx}
\usepackage{dcolumn}
\usepackage{bm}
\usepackage{color}
\usepackage{enumitem,kantlipsum}
\usepackage{bm}
\usepackage[normalem]{ulem}
\usepackage{color}
\numberwithin{equation}{section}
\usepackage{float}
\restylefloat{table}

\usepackage{lipsum}
\usepackage{gensymb}

\newcommand{\bq}{\mathbf{q}}

\newcommand{\br}{\mathbf{r}}

\newcommand{\bu}{\mathbf{u}}

\newcommand{\bk}{\mathbf{k}}

\newcommand{\beq}{\begin{equation}}
\newcommand{\eeq}{\end{equation}}
\newcommand{\beqn}{\begin{eqnarray}}
\newcommand{\eeqn}{\end{eqnarray}}

\newcommand{\dd}{{\rm d}}

\newcommand{\cO}{{\cal O}}

\newcommand{\lcn}{{L_H}}

\newcommand{\kbt}{k_BT}
\newcommand{\ko}{\kappa_0 }

\begin{document} 
\tighten
\newcommand {\be}{\begin{equation}}

\newcommand {\ee}{\end{equation}}

\newcommand {\gvt}{{\gamma v_2}}

\newcommand {\Dp}{{D_\perp(\bq)}}

\newcommand {\Dpr}{{D_{\rho\perp}(\bq)}}

\newcommand {\xpa}{the direction of mean flock motion $\hat{x}_{_\parallel}$}

\newcommand {\bea}{\begin{eqnarray}}

\newcommand {\eea}{\end{eqnarray}}

\newcommand {\tq}{\theta_{{\bf q}}}

\newcommand {\bqp}{{\bf q}_\perp}

\newcommand {\xpl}{{\hat{\bf x}_{_\parallel}}}

\newcommand{\vt}{{\bf v}_{_T}}

\newcommand {\vq}{\overline{|{\bf v}_{_\perp}({\bf q})|^2}}

\newcommand {\rsv}{\overline{|{\bf v}_{_\perp}({\bf r})|^2}}

\newcommand {\red}{\texttt{\color{red}}}
\newcommand {\blue}{\texttt{\color{blue}}}
\newcommand{\vp}{{\bf v}_{_\perp}}
\newcommand{\qp}{{\bf q}_{_\perp}}
\newcommand{\chip}{ \xi_{_\perp}}
\newcommand{\chipar}{ \xi_{_\parallel}}
\newcommand{\rpar}{ r_{_\parallel}}

\newcommand{\qpar}{ q_{_\parallel}}
\newcommand{\mqp}{ q_{_\perp}}

\newcommand {\fl}{\cite{TT1, TT2,TT3,TT4, NL}}

\def\lsim{\:\raisebox{-0.5ex}{$\stackrel{\textstyle<}{\sim}$}\:}

\def\gsim{\:\raisebox{-0.5ex}{$\stackrel{\textstyle>}{\sim}$}\:}

\def\Dtens{\mbox{\sffamily\bfseries D}}

\def\Wtens{\mbox{\sffamily\bfseries W}}

\def\Ptens{\mbox{\sffamily\bfseries P}}

\def\Otens{\mbox{\sffamily\bfseries O}}

\def\TT{\cite{TT1,TT2,TT3,TT4,NL}}

\def\Qtens{\mbox{\sffamily\bfseries Q}}

\def\Q{\mbox{\sffamily\bfseries Q}}

\def\Ntens{\mbox{\sffamily\bfseries N}}

\def\Ctens{\mbox{\sffamily\bfseries C}}

\def\Itens{\mbox{\sffamily\bfseries I}}

\def\Atens{\mbox{\sffamily\bfseries A}}

\def\A{\mbox{\sffamily\bfseries A}}

\def\Ktens{\mbox{\sffamily\bfseries K}}

\def\Vtens{\mbox{\sffamily\bfseries V}}

\def\Gtens{\mbox{\sffamily\bfseries G}}

\def\ftens{\mbox{\sffamily\bfseries f}}

\def\vtens{\mbox{\sffamily\bfseries v}}

\def\nabbold{\mbox{\boldmath $\nabla$\unboldmath}}

\def\nabvec{\mbox{\boldmath $\nabla$}}

\def\sigtens{\mbox{\boldmath $\sigma$\unboldmath}}

\def\etatens{\mbox{\boldmath $\eta$\unboldmath}}

\def\beq{\begin{equation}}

\def\bea{\begin{eqnarray}}

\def\eeq{\end{equation}}

\def\eea{\end{eqnarray}}

\def\etabold{\mbox{\boldmath $\eta$\unboldmath}}
\def\beq{\begin{equation}}
\def\eeq{\end{equation}}
\def\bea{\begin{eqnarray}}
\def\eea{\end{eqnarray}}
 \title{Statistical mechanics of asymmetric tethered membranes: spiral and crumpled phases }
\author{Tirthankar Banerjee}\email{tirthankar.banerjee@u-psud.fr}
\affiliation{LPTMS, UMR 8626, CNRS, Univ. Paris-Sud, Universit\'e Paris-Scalay, 91405 Orsay Cedex, France}
\affiliation{Condensed Matter Physics Division, Saha Institute of
Nuclear Physics, Calcutta 700064, West Bengal, India}
\author{Niladri Sarkar}\email{niladri2002in@gmail.com}
\affiliation{Max-Planck Institut f\"ur Physik Komplexer Systeme, N\"othnitzer 
Str. 38,
01187 Dresden, Germany}\affiliation{Laboratoire Physico
Chimie Curie, UMR 168, Institut Curie, PSL Research University,
CNRS, 
Sorbonne Universiti\'e,
 75005 Paris, France.}
\author{John Toner}\email{jjt@uoregon.edu}
\affiliation{Department of Physics and Institute of Theoretical Science, University of Oregon, Eugene, Oregon 97403, USA}
\author{Abhik Basu}\email{abhik.basu@saha.ac.in,abhik.123@gmail.com}
\affiliation{Condensed Matter Physics Division, Saha Institute of
Nuclear Physics, Calcutta 700064, India} 
\affiliation{Max-Planck Institut f\"ur Physik Komplexer Systeme, N\"othnitzer 
Str. 38,
01187 Dresden, Germany}

\date{\today}
\begin{abstract}
  We  develop the elastic theory for inversion-asymmetric tethered 
membranes and use it to identify and study their possible phases.  Asymmetry in a tethered membrane causes spontaneous curvature, which in general depends upon the 
local in-plane dilation of the tethered network. This in turn leads to  long-ranged interactions between the local mean and Gaussian curvatures,  which is not present in symmetric tethered membranes.   This interplay between asymmetry and Gaussian curvature leads to  a new 
{\em double-spiral} phase
not found in symmetric tethered membranes. At temperature $T=0$,
 tethered membranes
of arbitrarily large size are always rolled 
up 
tightly  into a conjoined pair of Archimedes' spirals. At finite $T$  this spiral  structure swells up significantly  into algebraic spirals characterized by universal exponents which we calculate.  
These spirals have long range orientational order, and are  the asymmetric analogs of 
statistically flat symmetric tethered membranes.   We also find that sufficiently strong asymmetry can trigger a structural instability 
leading to crumpling of these membranes as well. This provides a new 
route to crumpling  for asymmetric  tethered membranes. We  calculate the maximum linear extent  $L_c$  beyond which the membrane crumples, and calculate the universal dependence of $L_c$ on the membrane parameters. By tuning the asymmetry parameter, $L_c$ can be continuously varied, implying a {\em scale-dependent} crumpling.
Our theory can be tested on controlled experiments on lipids with artificial deposits of spectrin filaments,
 in-vitro experiments on 
red blood cell membrane extracts, 
and on graphene coated on one side. 
\end{abstract}

\maketitle

\section{Introduction}\label{intro}

The statistical mechanics of membranes has  
  long generated considerable 
theoretical and experimental interest~\cite{nelson2004}. 
In contrast to linear polymers~\cite{cates1984,cates1985}, fluctuating surfaces 
 can exhibit a wide variety of different phases, 
depending on rigidity, surface tension, and 
various microscopic constraints.
Polymerized or tethered membranes,
are particularly interesting~\cite{nelson2004,kantor1987}.
These are 
two-dimensional (2D) analogs of linear polymer chains. But, unlike polymers, 
which are always coiled up, 
tethered membranes  at low temperatures ($T$) or high bending 
rigidity are known~\cite{nelson2004,paczuski1998} to 
display a statistically flat phase with long range orientational order in 
the surface normals. Notice 
that the very existence of a 
2D flat phase is surprising, since the well-known Hohenberg-Mermin-Wagner  (HMW)
theorem  forbids spontaneous symmetry breaking for two 
dimensional systems with a continuous symmetry~\cite{mermin1966,hohenberg1967}. 
This apparent 
violation of the HMW theorem is 
possible due to the coupling between 
the in-plane elastic degrees of freedom and the out-of-plane undulations,
 which
introduces an effective long-ranged interaction between the undulation modes. 
 Since the HMW theorem only applies for systems with short-ranged interactions, 
this long-ranged interaction allows tethered membranes to have the long-ranged 
orientational order that must occur in a flat phase.
At higher temperature, 
tethered membranes 
possibly show a phase transition to a {\em crumpled} 
phase~\cite{nelson2004,guitter1989,david1996,wiese1997}, although the 
existence of the latter remains controversial 
 even now~\cite{domb2000}.

Most  theoretical studies of tethered membranes to 
date  that we know 
of 
 have considered only
 {\it inversion-symmetric} membranes, i.e.,  
membranes that are identical 
on both sides. 
 Many real membranes, e.g., graphene coated on one side by some 
substance (e.g., polymer or a layer of lipid) and both {\em 
in-vivo} red blood cell membranes and {\em in-vitro} spectrin-deposited model 
lipid bilayers~\cite{lopez2012} are {\em structurally}  {\em inversion 
asymmetric}.
The effects of {\it such} asymmetry 
 are  still  largely unexplored  theoretically.

\begin{figure}[htb]  
 \includegraphics[width=6cm]{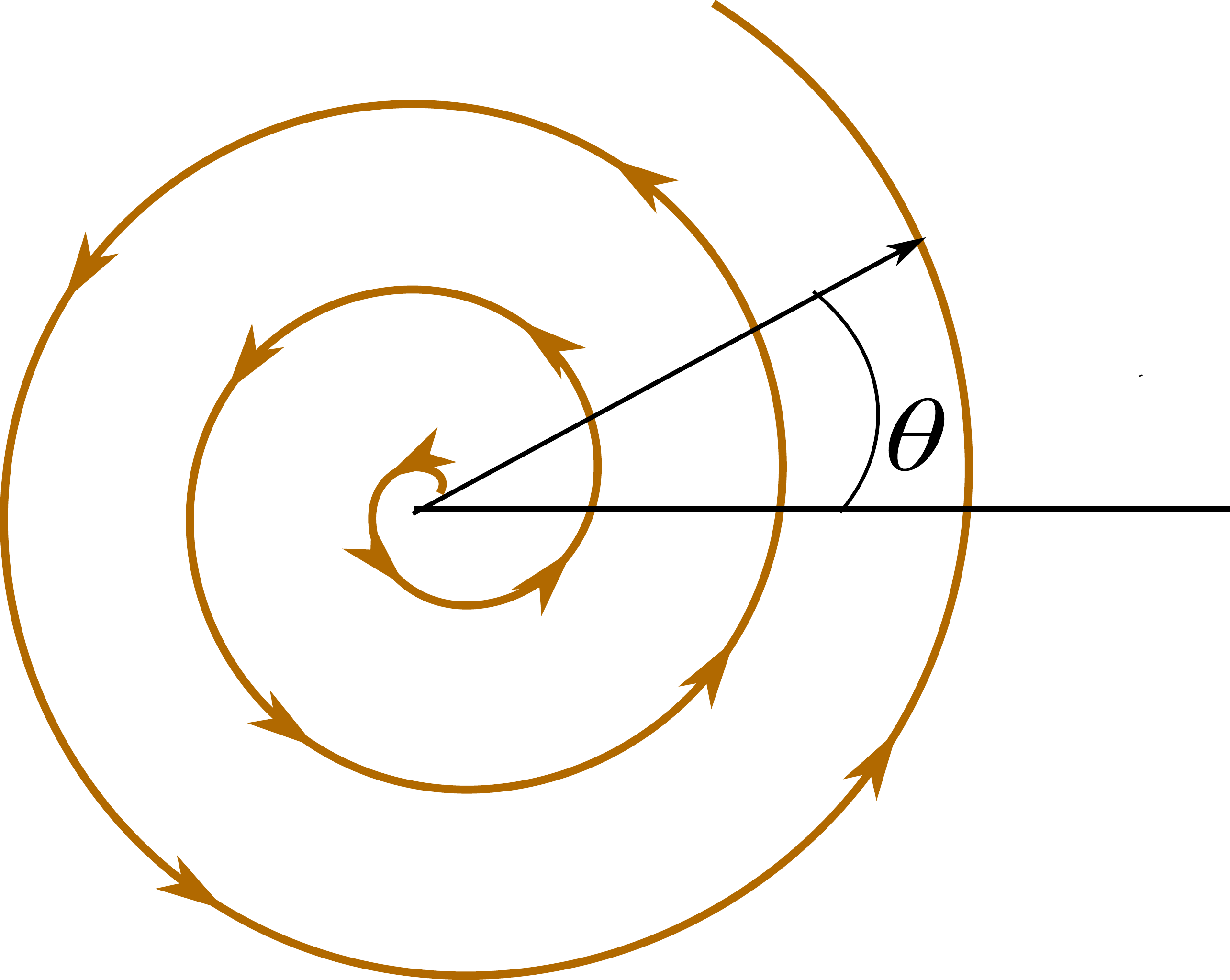}\vspace{0.8cm}
 \includegraphics[width=6cm]{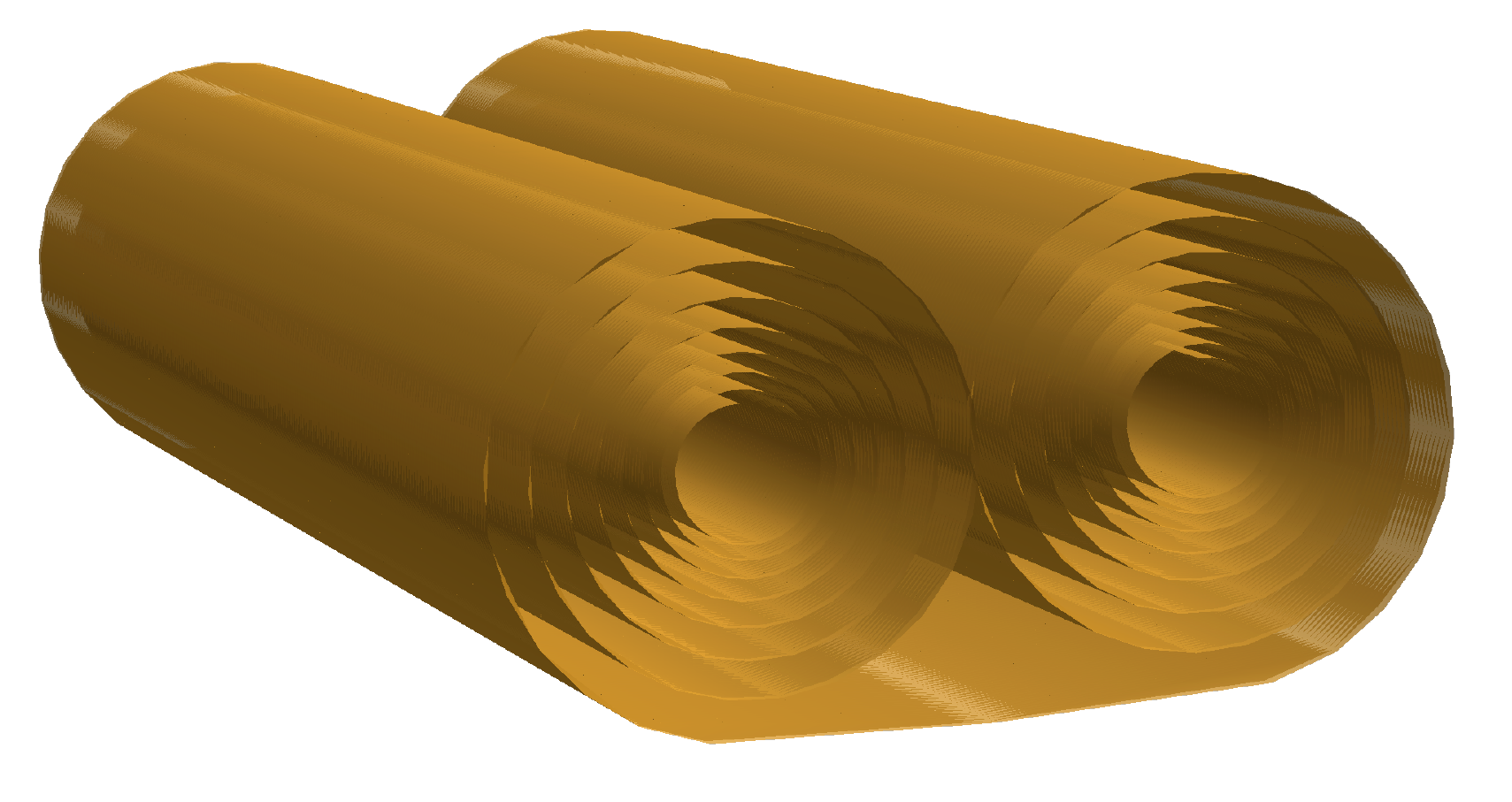}
 \caption{(Color online)(top) Schematic diagram of the cross section of a 
 spiral, (bottom) Schematic diagram of the double spiral structure of 
 our model membrane. } 
\label{dspiral}
 \end{figure}

In this paper, we 
 develop a generic and experimentally testable theory of 
equilibrium 
asymmetric tethered membranes. 
We find that such membranes exhibit a new 
"spiral state" not found in symmetric membranes.  As illustrated in 
Fig.~\ref{dspiral}, 
the mean spatial configuration of this state can be obtained by joining two 
coplanar spirals of opposite handedness at their base, and extruding that curve 
in the direction perpendicular to the plane of the spirals.

The shape of the spirals in the spiral state is universal.
First consider membranes  
for which thermal fluctuations are negligible  (i.e., membranes that are effectively at temperature $T=0$).  Such asymmetric membranes  of 
large linear size $L_m$
 spontaneously arrange themselves into a double spiral of Archimedes\cite{arch} 
 structure:~\footnote{Unsurprisingly, inversion-symmetric 
tethered membranes are always uncrumpled and flat at temperature 
$T=0$.}
\bea
r(\theta)=r_0 + a{\theta \over 2\pi}, \label{rspiral}
\eea
 where $r(\theta)$ is the radius of the spiral from its center to a point on 
the spiral 
at which the radius is being determined and $\theta$ is the angle of that 
point in the 
anticlockwise direction as shown in Fig.~\ref{dspiral} (left) for the 
right-hand spiral, 
and in the clockwise direction for the left hand spiral, with $\theta=0$ 
 being the innermost edge of the membrane;  see Fig.~\ref{dspiral} (right) for 
a 
schematic 
picture of a double spiral.
 In (\ref{rspiral}), $a$ is the 
thickness of the membrane. Choosing this form 
for $r(\theta)$ simply means that the membrane is curled up as tightly as it can, 
given excluded volume effects. In (\ref{rspiral}), the size $r_0$ of 
the 
{\em hole} 
left in the center of the spiral is given by 
\bea
r_0={\kappa_0 \over 2C}. \label{r011} 
\eea
 where $\kappa_0$ is  the "bare" bend modulus  of the membrane (to be defined formally below)  and $C$ is a 
phenomenological 
"spontaneous curvature" parameter which is a measure of the asymmetry of the 
membrane, and 
 will  also be defined more precisely below. Since $r_0$ is independent of the size 
of 
the membrane, it is always negligible compared to the outer radius of the spiral 
for a sufficiently large membranes ($L_m\gg r_0$). Thus, one can effectively 
consider the spiral 
to extend all the way into the origin.
 
 Thermal fluctuations considerably change this picture.   
 For membranes with sufficiently small asymmetry,  thermal fluctuations open up  
the spiral
  by giving rise to a longer ranged "Helfrich repulsion"\cite{helfrich1978}; the resultant form of the the spiral is: 
\bea
r(\theta)=R_0\theta^\nu, \label{rtheta} 
\eea 
where the universal exponent $\nu$ is related to the equally universal 
exponent $\eta$ characterizing the anomalous bend elasticity~\cite{nelson2004} of 
{\it symmetric} membranes through the relation 
\beq
\nu={4 \over 2+\eta}{ 
\approx4-{2\over3}\sqrt{15}\approx1.418} \,,
\label{nu}
\eeq
where the numerical estimate is 
based on the theoretical estimate
\bea
\eta\approx{4\over 1+\sqrt{15}}\approx .821
\label{eta}
\eea
 obtained by  Le Doussal and Radzihovsky~\cite{Leo}. For a large enough 
spiral, consecutive segments of size smaller than $\kappa/C$ appear nearly 
flat, and hence behave like a stack of symmetric membranes locally.
 In addition, the scale length $R_0$ exhibits universal scaling with 
temperature and other parameters, which can also be related exactly to the 
exponent $\eta$; we find
\bea
R_0&=&\left[(k_B T)^{2(2-\eta)}\kappa_0^{2(\eta-1)}A_0^{-{\eta}}C^{(\eta-2)}\right]^{ 1/(2+\eta)}\times\cO(1) \,,\nonumber\\ &=& (k_B T)^{.836}\kappa_0^{-.127}A_0^{-.291}C^{-.418}\times\cO(1) \,,
\label{R_0}
\eea
where $\kappa_0$ is  the "bare" bend modulus  (to be defined formally below)  and 
$A_0\equiv\frac{4\mu_0(\mu_0+\lambda_0)}{2\mu_0+\lambda_0}>0$, with $\mu_0$ and 
$\lambda_0$  the equally bare  two-dimensional Lame' elastic coefficients of the membrane.  Here, by ``bare", we mean the values these parameters have 
before being renormalized by thermal fluctuation effects.  The numerical
values for the exponents quoted in the second line are based on the estimate (\ref{eta}) of $\eta$.

 All of the parameters in this paper, and the equations defining them, are summarized in the glossary that constitutes appendix (\ref{gloss}) of this paper.

The total radius $R_T$ of the spiral regions also exhibits universal scaling,  in 
this case  with
the spatial extent $L_m$ of the membrane:
\bea
R_T= R_0^{1-\alpha} L_m^\alpha \,\,\,\,,\,\,\,\, \alpha\equiv{4\over6+\eta}\approx 0.586\,\,.
\label{ralpha} 
\eea

The entire picture of the spiral state just described presupposes that each of the two spirals makes many turns. It therefore behooves us to ask how many turns $n$ the spirals formed by a membrane of length $L_m$ actually make. Assuming this number is large, it is easily found by plugging the total radius $R$ obtained from (\ref{ralpha}) into our expression (\ref{rtheta}) for the spiral structure, equating $\theta$ on the right hand side of that expression to $2\pi n$, and solving for $n$. We thereby obtain 
\beq
n(L_m)={1\over2\pi}\left({L_m\over R_0}\right)^{\alpha/\nu}={1\over2\pi}\left({L_m\over R_0}\right)^{\varpi}\,,
\label{turns}
\eeq
where we have defined another universal exponent 
\beq
\varpi\equiv{\alpha\over\nu}={2+\eta\over6+\eta}\approx 0.414 \,.
\label{nexp}
\eeq

Note that, based on our earlier expression for the length scale $R_0$, and the numerical estimate (\ref{eta}) of $\eta$, the number of turns is quite insensitive to material parameters 
\beq
n(L_m)\propto  R_0^{-\varpi}\propto T^{-0.346}\kappa_0^{0.053}A_0^{0.12}C^{0.173}\,,
\label{turnscale}
\eeq
so we can estimate the number of turns fairly accurately even if there is a large uncertainty in the values of the material parameters. For example, consider a graphene sheet made asymmetrical by being coated with cholesterol. The Young's modulus of graphene is~\cite{graphenemu}
$G=10^{12}$ Pa. If we model graphene as a bulk elastic sheet with this Young's modulus and thickness~\cite{graphenethick} $a=3.7
\AA$ (the interatomic distance), then  we can estimate $\kappa_{ 0}\sim Ga^3=5\times10^{-10}$ergs and
 $A_0\sim Ga=3.7{ \times10^5{{\rm dynes}\over {\rm cm}}}$.   The parameter $C$ is trickier to estimate; if we assume that it is 
 comparable to its value for pure cholesterol, and estimate that value by using  the value of ${\kappa\over 
 C}=2.9\times10^{-9}{\rm m}$ for pure cholesterol (see table~\ref{tab2}), and estimating $\kappa_0$ for pure c
 holesterol by the value of $\kappa_0=4\times10^{-12}{\rm ergs}$ for DMPC (see table~\ref{tab2}), we get 
 $C=1.38\times10^{-10}{\rm N}$. Using these values in blah for $R_0$ gives $R_0=2.54\times10^{-12}{
 \rm m}$; using that in (\ref{turns}) gives $n=27\times\left({L_m\over1\mu}\right)^\varpi=27\times\left({L_m\over1\mu}\right)^{ 0.414}$. 
 
 One could very well question all of the above estimates  of the material parameters $C$, $\kappa_0$, and $A_0$; but, due to the insensitivity of $n$ to  those parameters, the estimate will not change very much: any membrane larger than about one micron   should exhibit a sufficient
number of turns for our theory to be valid.

This spiral  state is 
not the only possible phase of an 
asymmetric membrane: we also find that asymmetric tethered membranes exhibit a 
crumpled 
phase.    Indeed, we have discovered 
a novel structural instability in
asymmetric 
membranes, in which
asymmetry  actually {\it induces} 
crumpling of the membrane~\cite{peliti1985,nelson2004}  
More specifically, we find that   sufficiently asymmetric tethered membranes in 
equilibrium become {\em structurally unstable }, yielding a  {\em crumpled 
state for sufficiently large  asymmetry.}

This instability is driven by the dependence of the local spontaneous curvature on the local dilation of the tethered network, a dependence that on symmetry grounds can only occur in asymmetric membranes. The strength of this dependence is given by a dilation-bend elastic coupling constant $\chi$ that can be used as a measure of the degree of asymmetry, since it is only non-zero in asymmetric membranes. The critical value of $\chi$ at which this effect of asymmetry drives crumpling is determined
 by the "decoupled" bend modulus $\kappa'$
 where by ``decoupled", we mean the bend modulus the membrane would have in the absence of the  dilation-bend elastic coupling constant $\chi$. The parameters $\kappa'$ and  $\chi$ are defined precisely in equation (\ref{free energy}) below.

\begin{figure}[htb]  
 \includegraphics[width=7cm]{phase2.pdf}\hfill
\includegraphics[width=8cm]{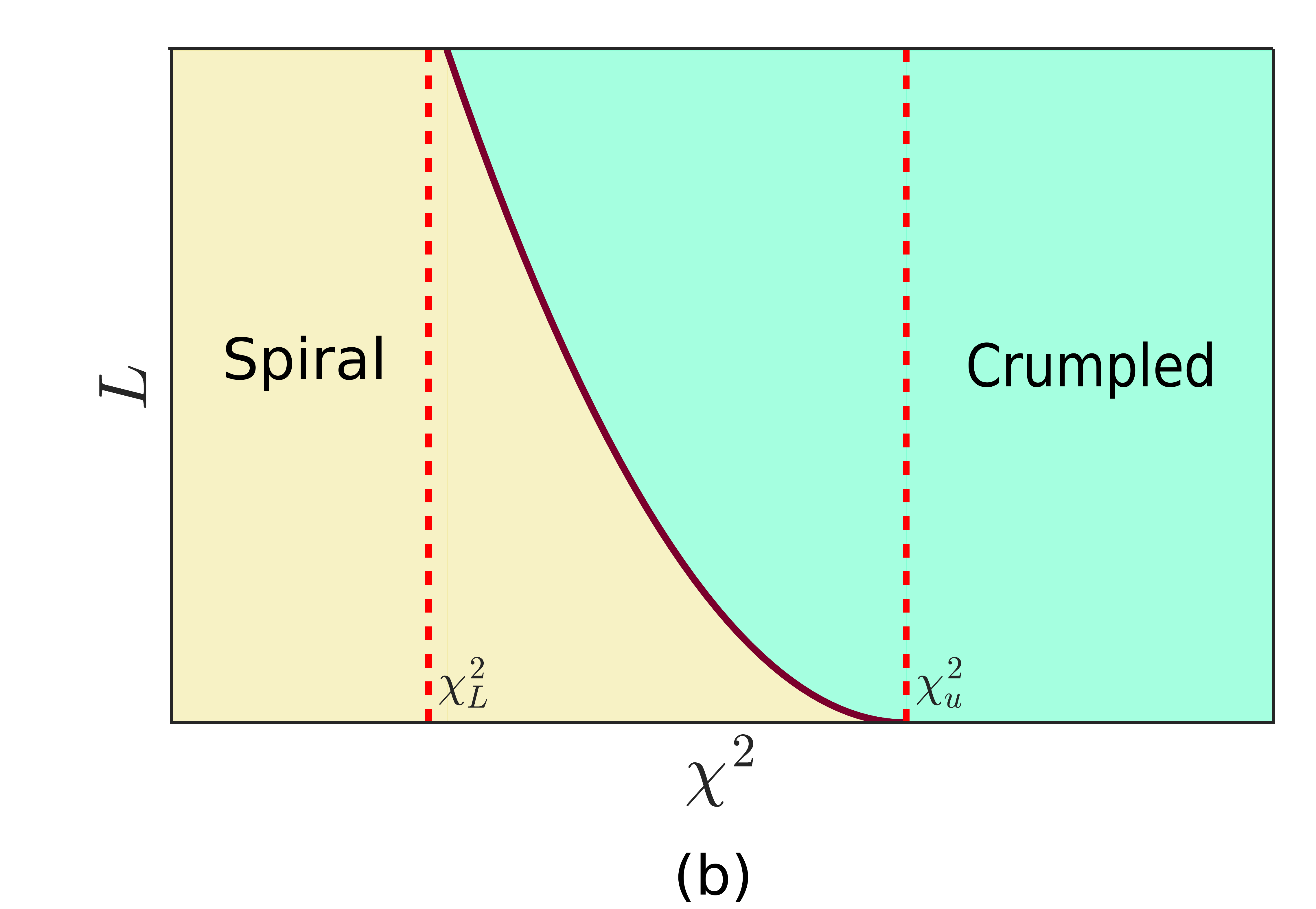}
 \caption{(Color online) 
 a)   Schematic 
phase diagram in the $\chi^2-\kappa'$ plane. 
\\
b)The 
continuous curve (black) is the line $L=\xi (\chi^2)$, demarcating the  
spiral
and the crumpled phases.}
\label{k-chi}
 \end{figure}

 The phase diagram for an asymmetric membrane in the $\chi$-$\kappa^\prime$ 
 plane is illustrated in Fig.~\ref{k-chi}a.  To connect this diagram to experiments, we note that in general both $\kappa'$ and $\chi$ should be functions of almost every imaginable experimental control parameter; e.g., temperature and salt concentration in the fluid around the membrane. If this function is analytic, which we expect it to be in general, then the topology of the phase diagram plotted as a function of any two experimental control parameters (e.g., temperature and salt concentration) will  have the same topology as Fig.~\ref{k-chi}. 

Note that this phase diagram has two distinct phase boundaries.  The lower of these, $\chi_{_{_L}}(\kappa')$, separates 
 two distinct regimes of parameter space  within this crumpled phase.  In one  of these, (hereafter called the ``strongly crumpled" (SC) regime), 
 the membrane will be crumpled no matter how small it is, while in the second 
(hereafter called, "weakly crumpled" (WC)), it is only crumpled if 
its lateral spatial extent $L_m$ exceeds a critical size $L_c$. Smaller 
membranes (i.e., $L_m<L_c$) exhibit a spiral structure similar to that found in 
the spiral phase, but different in its scaling properties. This behavior is 
summarized in Fig.~\ref{k-chi}b.

If $\kappa_{ 0}  \gg\kbt$, 
most of  the boundary $L_c(\chi)$ between the crumpled and the spiral phases in Fig.~(\ref{k-chi})b obeys 
\beq 
L_c(\chi)\propto (\chi_{_U}^2 - \chi^2)^{7/2}  \,.\label{L_c11}
\eeq
This laws breaks down near the two limits 
$\chi\to\chi_{_{_U}}$, where $L_c$ gets to be $<R_0$, so no uncrumpled membrane can be long enough   to wind up into a spiral, and as $\chi\to\chi_{_{_L}}$, where 
$L_c$ diverges. Unfortunately, this divergence is controlled by a perturbatively inaccessible fixed point, as illustrated in figure (\ref{fp}), so we can say nothing quantitative about the functional dependence of $L_c(\chi)$ as $\chi\to\chi_{_{_L}}$.

\begin{figure}
 \includegraphics[width=8cm]{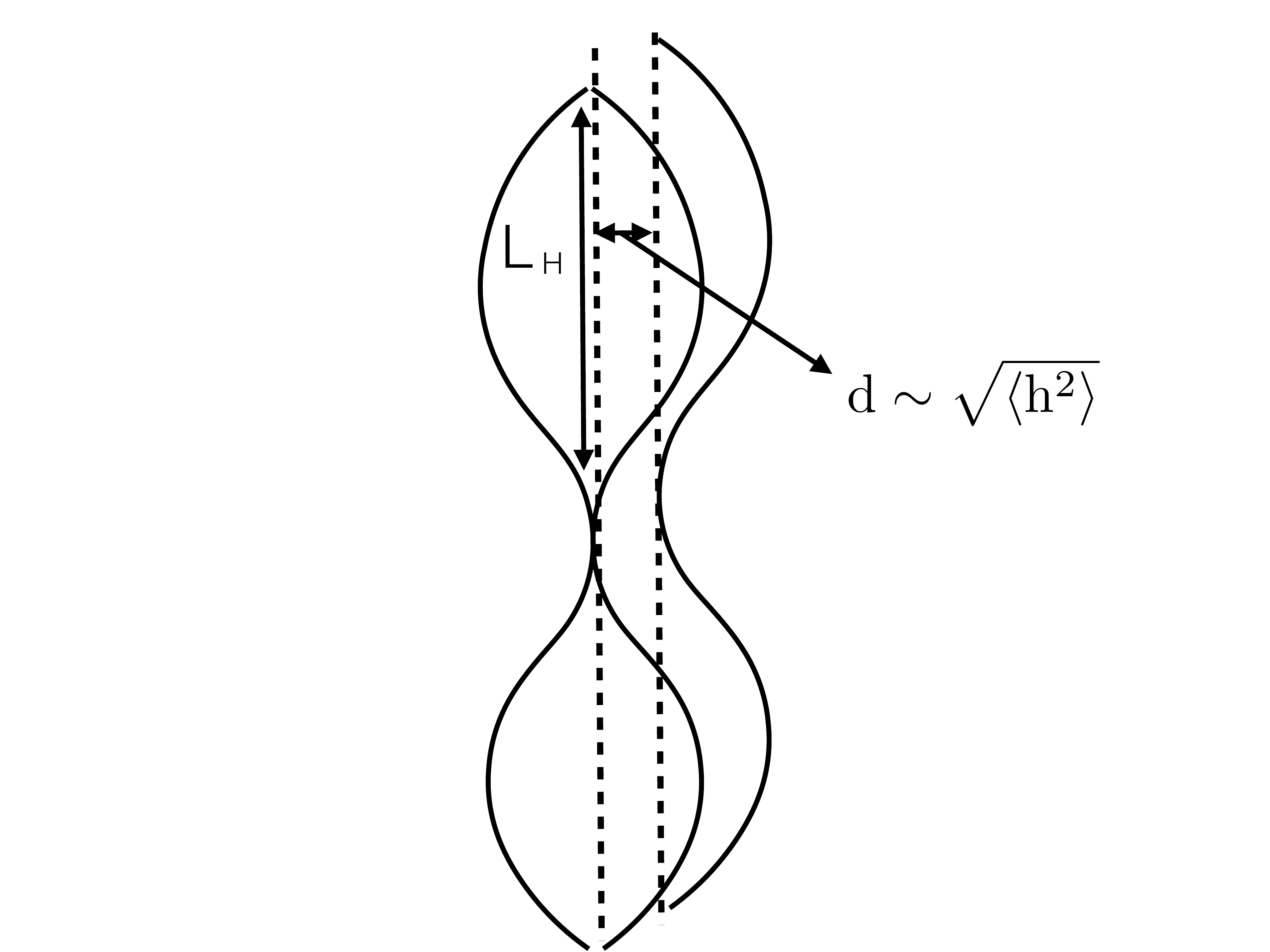}
 \caption{Schematic diagram depicting local patches of successive turns of spiral at finite $T$.The length scale $L_H$ is the typical distance between successive points of contact of neighboring layer, while $d\sim \sqrt{\langle h^2\rangle}$  is the typical distance between successive layers.  }
  \label{lengthscale}
\end{figure}

We   find that there are a hierarchy of length scales in asymmetric tethered membranes. While thermal fluctuations expand the spiral structure significantly from its $T=0$ shape, successive turns keep coming in contact with each other due to fluctuations. The smallest of the length scales in asymmetric tethered membranes  is the ``Helfrich length"
 $L_H$,  which is  the typical distance between ``bumps" or points of contact; see Fig.\ref{lengthscale}. 
For length scales $L\ll L_H$,
the asymmetric membrane looks flat, and self avoiding interactions are therefore unimportant\cite{nelson2004}.

 The 
remainder of this paper is organized as  follows. 
In section (\ref{L lt Lfree}), we formulate the elastic  theory, and study the behavior, of asymmetric membranes for the smallest range of length scales described above (i.e.,  $L \ll L_H$). In section (\ref{L gt Lfree}), we treat the largest length scales $L\gg L_H$, and determine the spiral structure, both for $T=0$, and $T\ne0$.  Section  (\ref{L gt Lfree}) also addresses crumpling, and demonstrates the existence of both the "weakly crumpled" and "strongly crumpled" regimes of parameter space, and the spiral  structure of membranes in the "weakly crumpled" regime that are small enough to avoid crumpling. In section (\ref{sum}), we summarize our results and discuss possible future theoretical and experimental work.



\section{Small length scales $L\ll L_H$}\label{L lt Lfree}

 In this section, we begin by formulating, in subsection (A), the elastic theory of asymmetric fluctuating membranes for length scales $L\ll L_H$, on which the membrane is  nearly flat and non-self intersecting. This differs from that for symmetric membranes by the addition of the two up-down symmetry breaking terms mentioned in the introduction:  the dilation-bend elastic coupling constant $\chi$, and the spontaneous curvature $C$. In subsection (B), we treat this model in the quadratic approximation, and demonstrate that sufficiently   strong asymmetry can cause crumpling. In subsection (C), we go beyond the harmonic approximation, and use the Renormalization group (RG) to treat the effect of elastic anharmonicities. We show that these are relevant, in RG sense of changing the long distance behavior of the membrane, if the internal dimension $d$ of the membrane is less than four (as it is in the physical case $d=2$).  We also find that for sufficiently small asymmetry, the membrane on length scales $L\ll L_H$ is controlled by the {\it same} RG fixed  point
as {\it symmetric} membranes; that is, the dilation-bend elastic coupling constant $\chi$
is irrelevant at long length scales, and the spontaneous curvature $C$, though relevant (i.e., growing upon renormalization), has not yet, at these short length scales, become important enough to matter.  For larger  asymmetry,  the RG flows do {\it not} approach the symmetric membrane fixed point; we argue that this implies the membrane crumples at these 
asymmetries, even for some asymmetries small enough that the harmonic theory would suggest that the membrane remains uncrumpled.

\subsection{Elastic free energy for $L\ll L_H$}\label{Fel}

We begin by 
formulating the elastic model for a single turn of the spiral structure, on 
length scales short compared to both the local radius of curvature $R$ and the 
typical distance $L_H$ between successive interactions of that turn with 
the turns
inside and outside of it: a membrane segment of linear size $L\ll L_H$ 
behaves 
like an isolated, free membrane not in contact with anything else. The results 
of this analysis will then be used in section (\ref{L gt Lfree}) as 
inputs to treat the membrane on progressively larger scales: first, to compute 
$L_H$, and thereby calculate the interaction between successive turns of 
the membrane, and then on length scales comparable to $R$, which will allow us 
to calculate the large scale spiral structure of the membrane.

On the smallest length scales $L$, we can ignore both self-avoidance 
interactions 
(since, by assumption, $L\ll L_H$) and the curvature of the membrane 
(since 
$L\ll R$). The latter simplification allows us to describe the membrane 
fluctuations in the so-called ``Monge gauge", which introduces  a single-valued  height field $h({\bf r})$  and an
 in-plane  displacement by a 2D vector field ${\bf u}({\bf 
r})$, 
${\bf r}=(x,y)$ with ${\bf R}(\br)\equiv ({\bf 
r}+\bu({\bf r}), h(\bf r)$) denoting the new, post-fluctuation coordinates in the 
three-dimensional embedding space
 of a point on the membrane  which was originally located at ${\bf R}(\br)=(x,y,z=0)$~\cite{nelson2004,chaikin2000}.

General symmetry considerations then 
dictate  the following  form 
 for the free energy functional ${\mathcal F}$ for 
tensionless asymmetric tethered membranes:
\begin{eqnarray}
{\mathcal F}&=&\frac{1}{2}\int d^2r \left[\kappa'(\nabla^2 h)^2+ 
\lambda u_{ii}^2 +
2\mu u_{ij}u_{ij} + 2\chi u_{ii}\nabla^2 h \right]\nonumber \\ &+&\int 
d^2r\,\,C\nabla^2 h
\label{free energy}
\end{eqnarray}
to 
leading order in gradients  (see Appendix \ref{full-rot} for a fully rotationally invariant free energy functional that yields (\ref{free energy}) in the nearly flat limit in the Monge gauge). Here $u_{ij}=\frac{1}{2}(\nabla_i u_j + 
\nabla_j u_i +\nabla_i h 
\nabla_j h)$ is the strain tensor, ignoring terms quadratic in $\nabla_i u_j$, 
which 
are irrelevant here in 
 the renormalization group (RG)
sense~\cite{chaikin2000,aronovitz1988,nelson2004}. 

  The free energy (\ref{free energy}),  differs from that of symmetric membranes only by the addition of   two generic 
inversion-symmetry breaking terms:  a linear
``spontaneous curvature" term $C\nabla^2 h$, that 
makes the membrane want to curl up
with a radius of 
curvature $R\propto 1/C$
and a  term  $\chi \nabla^2 h u_{ii}$, that describes local bending of the 
membrane in response to local compression of the elastic network. Both these
terms can be separately positive or negative, and  arise  
 naturally by expanding a local compression dependent curvature term 
$f(u_{ii})\nabla^2 h$ to 
linear order in $u_{ii}$. 
 A term analogous to our $\chi$ term was introduced for  {\it fluid} membranes by \cite{leibler1986}, a paper which inspired ours.

\subsection{Quadratic theory and zero-temperature asymmetry induced crumpling}\label{QT}

Up to 
  quadratic order in the fields, the free energy (\ref{free 
energy})  can be 
approximated as: 
 \begin{eqnarray}
  {\mathcal F}\approx {\mathcal F}_g &=&\sum_\bq 
 [\frac{\kappa'}{2}q^4 |h({\bf 
q})|^2 
  + (\frac{\lambda}{2}+\mu)q^2\{|u^{L}({\bf q})|^2\nonumber \\ &-& 
\frac{2iq\chi 
h({\bf q}) u^{L}({\bf -q})}{2\mu+\lambda}\}+\mu 
q^2|{\bf u}^{T}({\bf q})|^2 ],\label{bilinfree}
 \end{eqnarray}
where $h({\bf q})$, and ${\bf u}({\bf q})$ are the spatial 
Fourier transforms of
  $h({\bf r})$ and ${\bf u}({\bf r})$, 
 with
$u_i^L({\bf q})$ and $u_i^T({\bf q})$ the projections of ${\bf u}({\bf q})$ 
along and perpendicular to wavevector ${\bf q}$  
respectively. Note that $C$
has dropped out of the problem at this point; 
this is because the spontaneous curvature term, in the Monge approximation, is 
just a total derivative, and hence becomes a surface term which does not affect 
the Fourier modes. Once we go to  larger length scales  $L\gg L_H$ at which the Monge approximation 
breaks down due to spontaneous curvature of the membrane, this term will come 
into play; indeed, it will control the shape of the membrane, as we will see in section (\ref{L gt Lfree}) 
below.

 Integrating the 
 fields $u^T({\bf q})$ and $u^L({\bf q})$ 
  out of the   Gaussian (i.e., harmonic) approximation to the Boltzmann weight,    we obtain an effective free energy functional that depends only on $h({\bf q})$: 
\begin{widetext} 
 \beq
  {\mathcal 
P_g(\{h({\bf 
q})\}})=\int  {\mathcal D}u^T{\mathcal D}u^L{\exp(- \beta F_h(\{u^T({\bf q}), u^L({\bf q}), {h({\bf 
q})\}})\over {\mathcal Z_g}}\equiv {\exp(- \beta F_{\rm g-eff}(\{ {h({\bf 
q})\}})\over {\mathcal Z_{\rm g-eff}}} \ ,
\label{Ph}
\eeq 
\end{widetext}
where $P_g(\{h({\bf 
q})\})$ is the Gaussian approximation to the probability distribution for  ${h({\bf 
q}})$,  and 
\beq
\label{Fgeff}
{\mathcal F}_{\rm g-eff} = \sum_\bq [\frac{\kappa_0}{2}q^4 
|h({\bf 
q})|^2] \ ,
\eeq 
with an 
effective   bend modulus
 $\kappa_0$  given by: 
 \begin{equation}\label{kappa}
  \kappa_{ 0}=\kappa'-\frac{\chi^2}{2\mu+\lambda} \, .
 \end{equation}
  Evidently, $\kappa_{0}<\kappa'$. 
Thermodynamic stability 
of the 
  membrane clearly requires $\kappa_0>0$, otherwise 
instability ensues.  Equation (\ref{kappa}) therefore implies with an instability threshold for $\chi$ given by 
  \beq
  \chi_{_U}^2=\kappa'(2\mu+\lambda)
  \label{chiu}
  \eeq
for all $q$.  Notice that the 
correction to $\kappa'$ in (\ref{kappa}) does not depend upon $T$ and hence 
the crumpling instability can take place even at $T=0$. That (\ref{kappa}) 
holds down to $T=0$ should not be surprising; Eq.~(\ref{kappa}) may also be 
obtained by minimizing ${\mathcal F}_g$   over
$\bu$ for fixed $h({\bf q})$. 

This  vanishing of $\kappa$ with increasing $\chi$ is the 
asymmetry-induced 
crumpling discussed earlier in the introduction. Since our result (\ref{kappa})  is  
$q$-independent,  membranes of 
{\em any  size, no matter how small} will be crumpled, provided $\kappa<0$,  which  we have just shown will happen for  $\chi^2>\chi_U^2$.
We will see in the next section that anharmonic effects actually cause the membrane to 
crumple for a larger range of $\chi$'s; specifically, when 
$\chi^2>\chi_{_{_L}}^2$, where $\chi_{_{_L}}^2<\chi_{_U}^2$. However, for $\chi^2$ in the intermediate range $\chi_{_{_L}}^2<\chi^2<\chi_{_U}^2$, 
crumpling only occurs if the membrane is sufficiently large.  This intermediate regime is the "weakly crumpled" region labelled ``WC" in figure (\ref{k-chi}), while the range $\chi^2>\chi_{_U}^2$, in which even arbitrarily small membranes crumple, 
 is the ``strongly crumpled" (``SC") region in that figure.
 \vspace{.3in}

\subsection{Anharmonic theory for $L\ll L_H$}\label{An}
\subsubsection{Eliminating in-plane displacements $\bu$}

 As in symmetric membranes\cite{aronovitz1988,chaikin2000,nelson2004}, 
anharmonic effects (particularly those arising from the $\nabla_i h \nabla_j h$ 
piece of $u_{ij}$)
substantially modify the behavior of  asymmetric membranes. 
  Here we treat these anharmonic effects using   a perturbative renormalization group (RG) analysis of 
the model (\ref{free energy}). 
 Before doing this, however,  it is first convenient to proceed just as we did in the harmonic theory, and integrate the in-plane displacement field $\bu$  out of the full {\it anharmonic} Boltzmann weight $\exp\left(-\beta F\right)$, where ${\mathcal F}$ is given by the full elastic energy (\ref{free energy}),  to obtain an effective free energy for $h$ alone.
Since ${\mathcal F}$
 is bilinear  in $\bu$,  even though it is anharmonic in $h$, we can do this integration {\it exactly}. 
 
 That is, we write
 \begin{widetext} 
 \beq
  {\mathcal 
P(\{h({\bf 
r})\}})=\int  {\mathcal D}\bu{\exp(- \beta F(\{{\bf u}({\bf r}), {h({\bf 
r})\}})\over {\mathcal Z}}\equiv {\exp(- \beta F_{\rm eff}(\{ {h({\bf 
r})\}})\over {\mathcal Z_{\rm eff}}} \ ,
\label{Phf}
\eeq 
\end{widetext}
where $P(\{h({\bf 
r})\})$ is the exact probability distribution for  ${h({\bf 
r}})$.

The integration over $\bu$ can now be done as follows:


 { Recall the definition of the symmetrized strain
 \begin{eqnarray}  
 u_{ij}(\br) = \frac{1}{2}(\partial_j u_i + \partial_i u_j+A_{ij}) \,, 
\label{straindef}
 \end{eqnarray}
 where we have defined
 \beq
 A_{ij}({\bf r})\equiv
(\nabla_i h) 
(\nabla_j h) \,.
\label{Adef}
\eeq We now consider the Fourier transform $A_{ij}(\bq)$ of $A_{ij}(\br)$, and 
use the fact that any 2D symmetric second rank tensor can be written 
as
a sum of transverse and longitudinal parts  to write: 
 \begin{equation}
  A_{ij}(\bq)=\frac{1}{2}\left[iq_i\theta_j(\bq)+iq_j\theta_i(\bq)+P_{ij}(\bq)\Phi(\bq)\right] \,,
  \label{Adecomp}
 \end{equation}
 where the `` transverse projection operator"
 \beq
 P_{ij}(\bq)\equiv\delta_{ij}-\frac{q_i q_j}{q^2}
 \label{Pdef}
 \eeq
projects 
any vector onto the space perpendicular to ${\bf q}$. Taking $P_{ij}$ times both 
sides of (\ref{Adecomp}), and summing over repeated indices $ij$ eliminates the $\theta$ terms, since by construction $P_{ij}q_i=P_{ij}q_j=0$ (i.e., the projection of $\bq$ perpendicular to itself is zero), and leaves an expression for $\Phi$:
\beq
\Phi=P_{ij}A_{ij} \,,
\label{Phiq}
\eeq
where we have used the fact that $P_{ij}P_{ij}=P_{ii}=1$, the 
last equality holding in $D=2$;  $\theta_i (\bq)$ is any vector.

Using our decomposition (\ref{Adecomp})  in the Fourier transform of our definition (\ref{straindef}) of the strain tensor, we obtain an expression for the Fourier transformed strain tensor:
\beq
u_{ij}(\bq)=\frac{1}{2}\left[iq_i \tilde u_j(\bq)
+iq_j \tilde u_i(\bq)+P_{ij}(\bq)\Phi(\bq)\right] \,,
\label{strainFT}
\eeq
where we have defined 
\beq
\tilde 
u_i(\bq)\equiv u_i(\bq)+\theta_i(\bq) \,.
\label{utildedef}
\eeq

Now rewriting the $\bu$-dependent terms in ${\mathcal F}$ (\ref{free energy}) in Fourier space, we have
\begin{eqnarray}
 \int d^2r  \, u_{ij} u_{ij}&=&\sum_\bq \frac{1}{4}\left[|q_i \tilde u_j(\bq)
+q_j \tilde u_i(\bq)|^2 + |P_{ij}(\bq)\Phi(\bq)|^2\right]\nonumber\\
&=& \sum_\bq \left[\frac{1}{2}\left(q^2 |\tilde \bu(\bq)|^2+|\bq\cdot\tilde\bu(\bq)|^2\right)+ \frac{1}{4}|\Phi(\bq)|^2\right] \,,\nonumber\\
\end{eqnarray}
where in the first equality we have again used the properties of the projection operator $P_{ij}$ to eliminate the cross terms between $\tilde \bu(\bq)$ and $\Phi(\bq)$
and in the second equality we have again used the fact that   $P_{ij}P_{ij}=P_{ii}=1$ in $D=2$.

Similar manipulations give
\beq
\int d^2 r \, u_{ii} u_{jj}= \sum_\bq  [|\bq\cdot\tilde\bu(\bq)|^2+ i \Phi(-\bq) 
\bq\cdot 
\bu(\bq) + \frac{1}{4}|\Phi(\bq)|^2] \,,
\label{lambdaFT}
\eeq
and
\beq
\int d^2 r \,  u_{ii}\nabla^2 h = -\sum_\bq  q^2\left[i\bq\cdot\tilde\bu(\bq)+  {1\over2}\Phi(\bq) \right]h(-\bq) \,.
\label{chiFT}
\eeq
Using these in our expression (\ref{free energy}) for the free energy ${\mathcal F}$, and, as we did for the harmonic approximation, breaking $\bu$ into its components $u^L$ along and $u^T$ perpendicular to wavevector $\bq$ respectively, we obtain:
\begin{widetext}
 \begin{eqnarray}
{\mathcal F}= \sum_\bq
&[&\frac{\kappa'}{2}q^4 |h({\bf 
q})|^2 
  + \left(\frac{\lambda}{2}+\mu\right)q^2|\tilde u^{L}({\bf q})|^2- 
{i\over2}[2q^3\chi 
h({\bf q})-\lambda q\Phi(\bq)] \tilde u^{L}(-\bq)+\mu 
q^2|\tilde u^{T}({\bf q})|^2 \nonumber\\
&-&{1\over2}\chi q^2\Phi(\bq)h(-\bq)+{1\over4}\left({\lambda\over2}+\mu\right)|\Phi(\bq)|^2]\,.
\label{bilinfree1}
 \end{eqnarray}
\end{widetext}

It is now completely straightforward perform the Gaussian integral in (\ref{Phf}). Note that integral can be rewritten 
\beq
\int  {\mathcal D}\bu=\int \prod_\bq du^L(\bq) \, du^T(\bq)=\int \prod_\bq d\tilde u^L(\bq) \, d\tilde u^T(\bq) \,,
\label{utoutilde}
\eeq
where the last equality holds since the Jacobian of the coordinate transformation (\ref{utildedef}) from $\bu(\bq)$ to $\tilde\bu(\bq)$ is unity, since it is simply addition of a constant, because $\theta(\bq)$ depends only on the height field $h$, which is constant for the purposes of the functional integral in (\ref{Phf}).

Doing these Gaussian integrals over $\tilde u^L(\bq)$ and $\tilde u^T(\bq)$ then gives
  \begin{eqnarray}\label{free energy_h_FT}
  {\mathcal F}_h &=& \frac{1}{2}\sum_\bq  [\kappa q^4 |h(\bq)|^2 +
  \frac{A}{4}\left|P_{ij}(\bq) A_{ij}(\bq)\right|^2 \nonumber \\
  &-& B q^2 h(-\bq) P_{ij}(\bq) A_{ij}(\bq)], \label{waveF}
 \end{eqnarray}
 where  we have defined the couplings 
 \beq
 A\equiv\frac{4\mu(\mu+\lambda)}{2\mu+\lambda}>0
 \label{Aelasdef}
 \eeq
  and 
\beq
B\equiv\frac{2\chi\mu}{2\mu+\lambda} \,,
\label{Belasdef}
\eeq
and we remind the reader that $A_{ij}$ is completely determined by $h(\br)$ via  (\ref{Adef}),  and the projection operator $P_{ij}$ is defined by (\ref{Pdef}).

This Fourier space expression is the one we will use in the next subsection for our RG analysis. As noted by \cite{nelson2004}, however, it is instructive, and will prove useful later in our analysis of the spiral state, to rewrite this expression in real space, where its connection to mean and Gaussian curvature becomes clear. In real space,  (\ref{free energy_h_FT}) becomes
  \begin{eqnarray}\label{free energy_h}
  {\mathcal F}_h &=& \frac{1}{2}\int d^2 r  [\kappa(\nabla^2 h)^2 +
  \frac{A}{4}(P_{ij}\nabla_i h \nabla_j h)^2 \nonumber \\
  &+& B (\nabla^2 h) (P_{ij} \nabla_i h \nabla_j h)+2C\nabla^2 h] \,,
 \end{eqnarray}
 where  we have restored the $C\nabla^2 h$ term~\footnote{In $\mathcal{F}_h$ we have ignored a nonlinear 
term of the form $C\nabla^2 h ({\boldsymbol\nabla}h)^2$ that would originate 
from the expansion of the area element $dS$ in Monge gauge, 
$dS=\sqrt{1+({\boldsymbol\nabla h})^2}=1+ ({\boldsymbol\nabla h})^2$ for small 
fluctuations. This has the critical dimension of $4$ and hence is formally as 
relevant as the $A$- and $B$-terms are. This will generate additional
corrections to $\kappa$, $A$ and $B$ at $O(C^2)$ or higher, in addition to
generating a correction to $C$ itself (which is $C \times O(1)$). 
Nonetheless, the stable fixed point structure of $g_1=2\epsilon/5$, $g_2 =0$ 
and $C=0$ still holds and all our results should work. In any case, the RG 
eigenvalue of $C$ as the coefficient of the nonlinear term is $(4-d-\eta)$, 
where as $C$ as the coefficient of the corresponding linear term has its RG 
eigenvalue $2-\eta$. Noting that $\eta=2\epsilon/5$, near $D=4$, $C$ as the 
coefficient of the linear term dominates over the corresponding nonlinear term 
for large length scales. Hence, this nonlinear term may be ignored.}.  Notice that in (\ref{free energy_h}) above we have written $\kappa$, rather 
than $\kappa_0$. This is because $\kappa$ will be renormalized at finite 
temperature away from its bare value $\kappa_0$ due to fluctuations. By $P_{ij}\nabla_i h \nabla_j h$ we simply mean the Fourier transform back to real space of $\Phi(\bq)=P_{ij}(\bq) A_{ij}(\bq)$. This 
 depends 
non-locally on the field $h$; specifically~\cite{nelson2004}, on the Gaussian curvature of the membrane. To see this, multiply both sides of our expression (\ref{Phiq}) for $\Phi(\bq)$ by $q^2$:
\beq
q^2\Phi(\bq)=q^2 A_{ii}-q_iq_jA_{ij} \,.
\label{GC1}
\eeq
Fourier transforming this back to real space gives
\begin{widetext}
   \begin{eqnarray}
   \label{GC2}
\nabla^2  \Phi(\br)&=&\nabla^2 A_{ii}-\nabla_i  \nabla_j A_{ij}=\nabla^2 |{\bf\nabla} h(\br)|^2-\nabla_i  \nabla_j  [(\nabla_i h )(
\nabla_j h) ] \, ,
 \end{eqnarray}
 \end{widetext}
 where in writing the second equality we have used the definition (\ref{Adef}) of $A_{ij}$ in real space.

 Expanding out the implied sums over repeated indices in this expression specifically in $D=2$ gives, after a little algebra and elementary calculus,
    \begin{eqnarray}
   \label{GC3}
\nabla^2  \Phi(\br)&=&2\left[(\partial_x^2h)(\partial_y^2h)-(\partial_x\partial_yh)^2\right]=2S(\br)\, ,
\nonumber\\
 \end{eqnarray}
where 
\beq
S(\br)\approx\det(\partial_x\partial_yh)|_\br={1\over R_1(\br)R_2(\br)}
\label{GCdef}
\eeq
is the local Gaussian curvature\cite{nelson2004} at $\br$}, with $R_{1,2}(\br)$ the two principle radii of curvature at $\br$~\footnote{This expression for $S({\bf r})$ is not exact, but is 
valid in the limit of nearly flat membrane, for which $|\nabla h|\ll1$.}. Thus, 
as first noted by \cite{nelson2004}, the $A$ term above represents a very strong, 
long-ranged interaction between Gaussian curvatures at different points on the 
membrane.  This leads to stiffening of symmetric tethered membranes, for 
which $B=0$ identically, 
that allows long-range orientational correlation to survive in the 
thermodynamic limit. The $B$ term, which is only allowed in the asymmetric 
case, likewise 
represents  a long-ranged interaction between Gaussian curvature and mean 
curvature.




Equation (\ref{GC3}) implies
 that $P_{ij}(\nabla_ih)(\nabla_jh)=\int d^2r^\prime V(|{\bf 
r}-{\bf r}^\prime|)S({\bf r}^\prime)$, where $S({\bf r})$ is the local Gaussian 
curvature at $\bf r$, and 
\beq
\label{Vmg}
V(|{\bf r}|)={1\over2\pi}\ln(r/a) 
\eeq 
is the inverse Fourier transform of 
$-1/q^2$ (or, equivalently, the solution of $\nabla^2V(|{\bf r}|)=\delta(|{\bf r}|)$), with $a$ an ultraviolet cutoff. Therefore,  
\begin{eqnarray}
&&\int d^2r \,(\nabla^2h)P_{ij}[(\nabla_i h)(\nabla_j h)]\nonumber \\ &=&\int d^2r \,
d^2r^\prime \,\nabla^2h({\bf r})V(|{\bf r}-{\bf r}^\prime|)S({\bf r}^\prime) \,.
\label{f_SM}
\end{eqnarray}
Likewise,
\begin{eqnarray}
&&\int d^2r \,\left(P_{ij}(\nabla_i h)(\nabla_j h)\right)^2\nonumber \\ &=&\int d^2r \,
d^2r^\prime \,S({\bf r})V_2(|{\bf r}-{\bf r}^\prime|)S({\bf r}^\prime) \,,
\label{f_SS}
\end{eqnarray}
where 
\beq
\label{Vgg}
V_2(|{\bf r}|)={r^2\over2\pi}\ln\left({r\over ae}\right) 
\eeq 
is the inverse Fourier transform of 
$1/q^4$(or, equivalently, the solution of $\nabla^4V_2(|{\bf r}|)=\delta(|{\bf r}|)$).

Using these results (\ref{f_SM}) and (\ref{f_SS}) in
(\ref{free energy_h}), we obtain 
\begin{widetext}
  \begin{eqnarray}
  {\mathcal F}_h &=& \frac{1}{2}\int d^2 r  \left(\kappa(\nabla^2 h)^2 +2C\nabla^2 h\right)
+\int d^2 r\int 
d^2r^\prime \,\left({A\over8}S({\bf r})V_2(|{\bf r}-{\bf r}^\prime|)S({\bf r}^\prime)+{B\over2}\nabla^2h({\bf r})V(|{\bf r}-{\bf r}^\prime|)S({\bf r}^\prime)\right) \,,\nonumber\\
\label{free energy_h_RS}
 \end{eqnarray}
\end{widetext}
with the long-ranged potentials $V(\br)$ and $V_2(\br)$ given by (\ref{Vmg}) and (\ref{Vgg}) respectively.

This shows that 
the $A$ term in the free energy ${\mathcal F}_h$    is a very strong long ranged interaction between Gaussian curvatures at different points $\br$, $\br^\prime$ on the membrane.

The $B$ term likewise is a long-ranged interaction between mean 
and Gaussian curvatures.  To see this, one need simply note that for nearly flat membranes, 
\beq
\nabla^2 h\approx{1\over R_1}+{1\over R_2}=M \,,
\label{mean curve}
\eeq
where $M(\br)$ is the mean curvature at $\br$.

 \subsubsection{Renormalization group analysis}
 \label{RGMT}
 
 We will now present the renormalization group (RG) analysis of the ``height only" free energy
${\mathcal F}_h$ given by
 (\ref{free energy_h}).

Since we will eventually perform this RG in an expansion around the critical internal membrane
dimension D=4, it is useful to generalize ${\mathcal F}_h$ to higher D than the physical case D=2. We will do this simply by considering the wavevector $\bq$ in  (\ref{waveF})
to have $D$ components. Note that this is a somewhat different analytic continuation to higher dimensions than that used by, e.g. Aronovitz and Lubensky~\cite{aronovitz1988}. Although, obviously, our results should extrapolate onto theirs (or vice-versa) in D=2, the two different continuations can, and do, lead to slight quantitative differences in other dimensions; in particular, near D=4.

\begin{widetext}
\begin{figure}[htb]
\includegraphics[width=9cm]{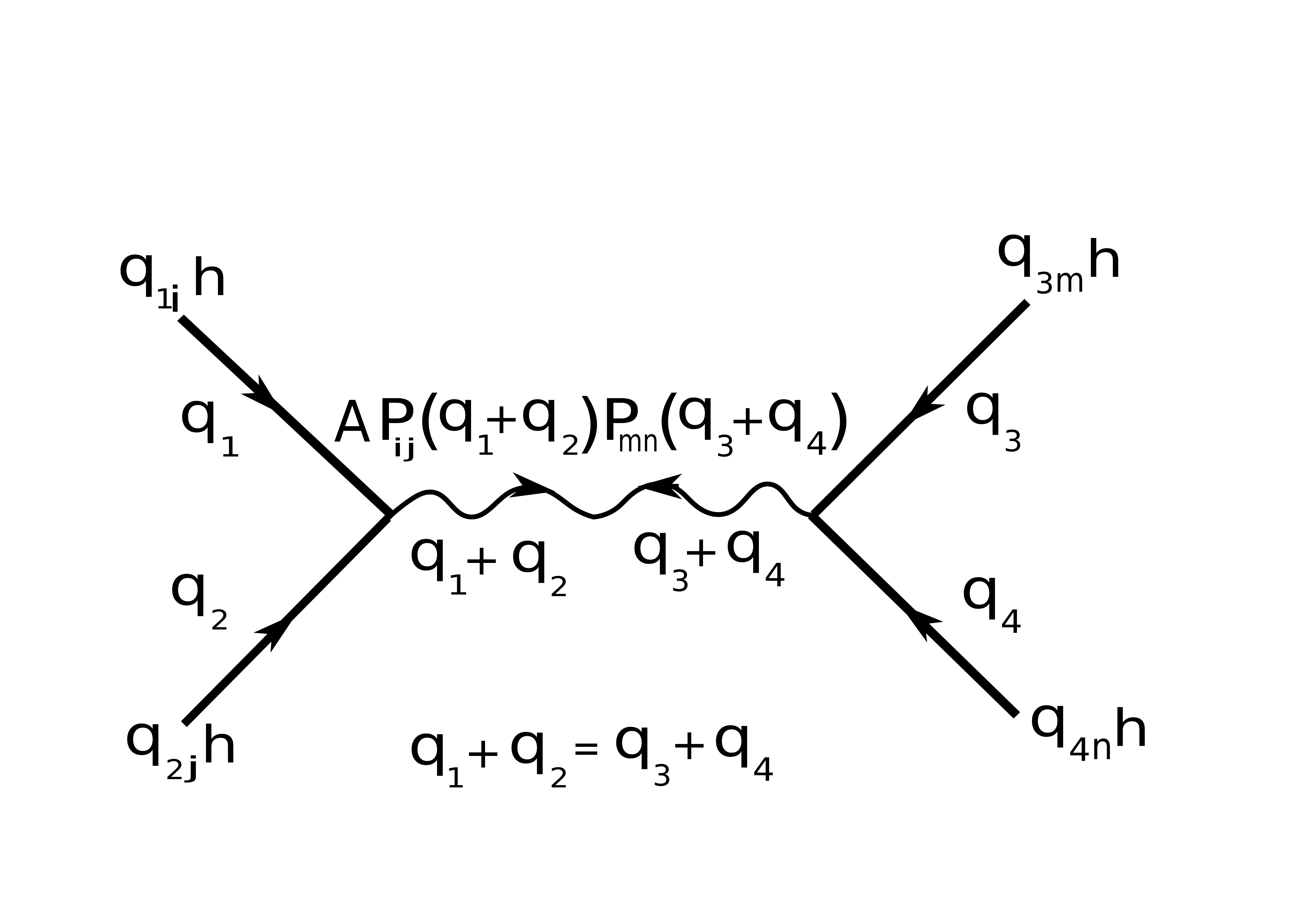}
\includegraphics[width=9cm]{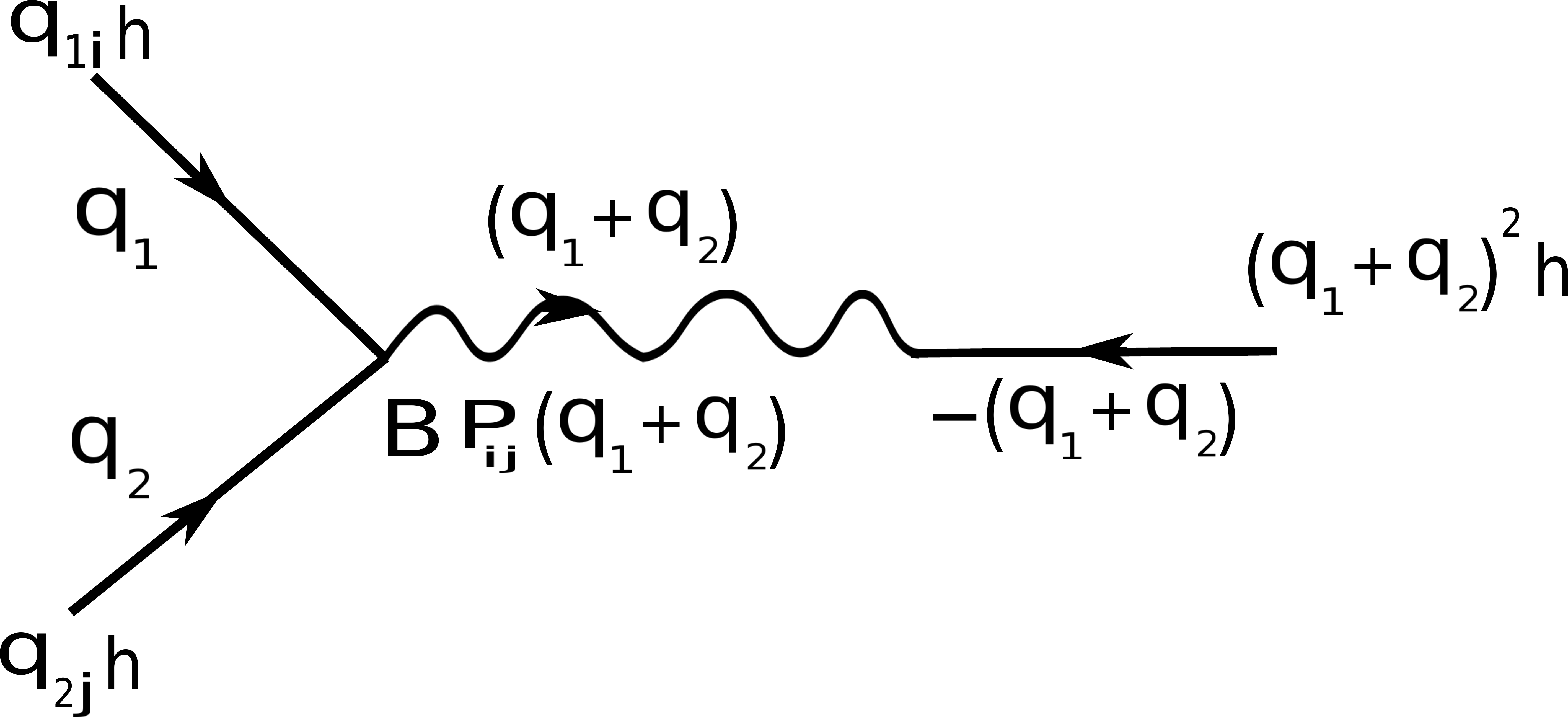}
 \caption{Vertices for the Feynman diagrams. 
 (left)$\frac{A}{4}(P_{ij}\nabla_i h \nabla_j h)^2$; 
 (right)$B (\nabla^2 h) (P_{ij} \nabla_i h \nabla_j h)$.
 }\label{vertices}
\end{figure}
\end{widetext}

The momentum shell RG procedure consists of tracing over the short wavelength 
Fourier modes
of $h(\vec{r})$, followed by a rescaling of lengths.
More precisely, we  follow the standard approach of initially restricting wavevectors  
to lie in a bounded spherical Brillouin zone: $|{\bf q}|<\Lambda$, where 
$\Lambda$ is an
ultra-violet cutoff,  presumably of order the inverse of the membrane thickness $a$  or spectrin mesh size, although its value has no effect on our  results. The height field ${\bf 
h}(\vec{r})$ 
is separated into high and low wave vector parts
$h(\vec{r})=h^<(\vec{r})+h^>(\vec{r})$,
where $h^<(\vec{r})$ has support in the large wave vector  (short wavelength) 
range $\Lambda
e^{-d\ell}<|{\bf q}|<\Lambda$, while $h^<(\vec{r})$ has support in the small 
wave vector (long wavelength) range $|{\bf q}|<e^{-d\ell}\Lambda$.
We then integrate out $h^<(\vec{r})$. This integration is done perturbatively 
in 
 the anharmonic couplings in (\ref{free energy_h}); as usual, this perturbation theory 
can be represented by Feynmann graphs, with the order of perturbation theory 
reflected by the number of loops in the graphs we consider.  
 The Feynman graphs (or ``vertices") representing the anharmonic couplings $\frac{A}{4}(P_{ij}\nabla_i h \nabla_j h)^2$ and $B (\nabla^2 h) (P_{ij} \nabla_i h \nabla_j h)$ are illustrated in 
Fig.~\ref{vertices}.

After this 
perturbative step, we rescale lengths, 
with ${\bf r}=r{\bf }' e^{\ell}$,  so as to restore the UV cutoff back to 
$\Lambda$.  This is then followed
by rescaling the long wave length part of the field $h({\bf 
q})=\zeta_h h^{\prime} ({\bf q}^{\prime})$;
 $\zeta_h=b^{(d+4-\eta)/2}$, $\eta$ being the anomalous dimension of $h$,  
which 
we will choose to produce fixed points. 

\begin{widetext}
\begin{figure}[htb]
\includegraphics[width=16cm]{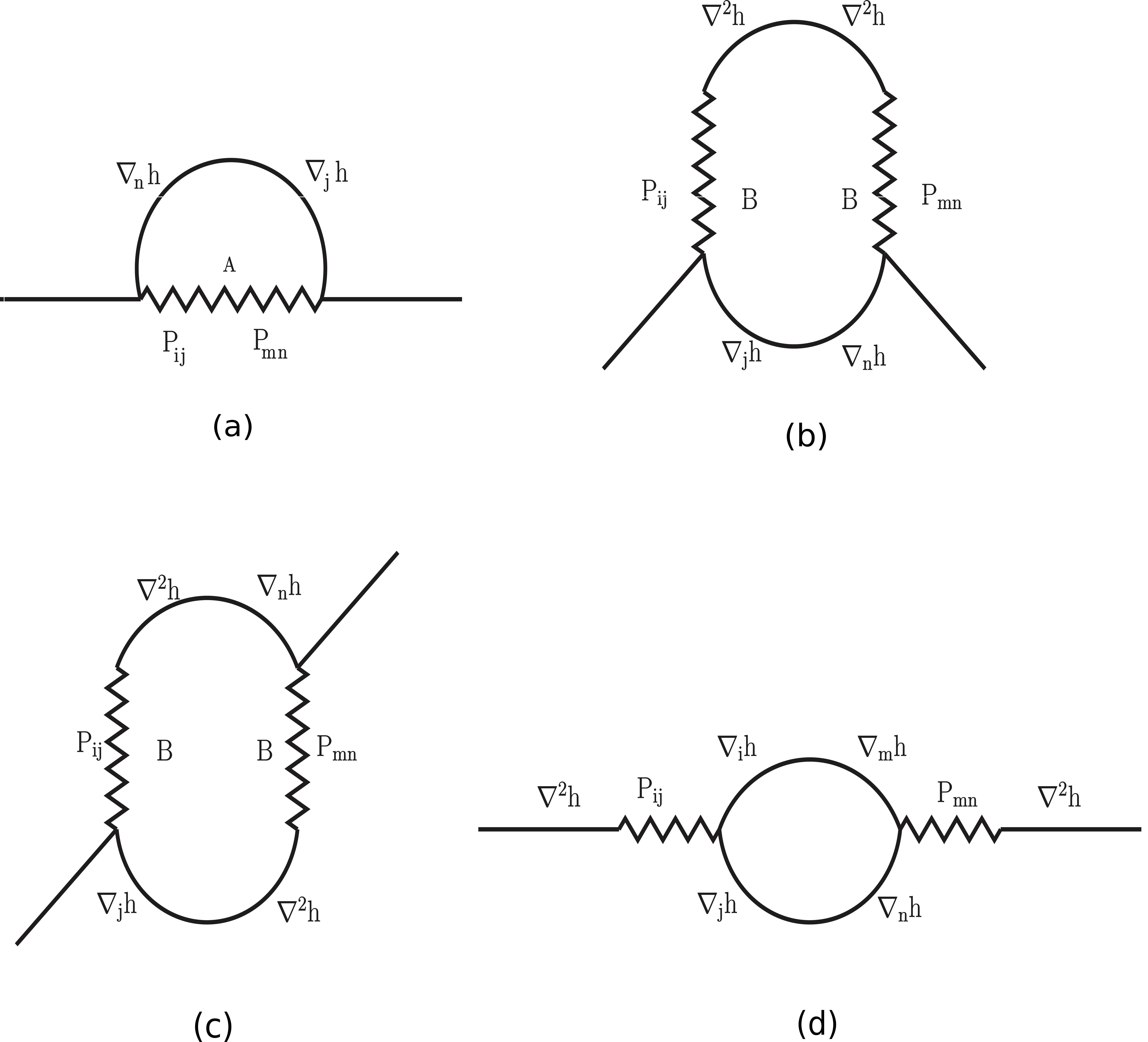}
 \caption{One-loop Feynman diagrams that contribute to the fluctuation 
corrections of $\kappa$.
}\label{kappa_dia-all}
\end{figure}
\end{widetext}

We restrict ourselves to a one-loop order renormalization group (RG) 
calculation.
At this order (equivalently, to the lowest orders in $A$ and $B$), $\kappa$ 
receives two fluctuation corrections, each 
originating from non-zero $A$ and $B$, respectively; the 
relevant Feynman diagrams are given in  Fig.~\ref{kappa_dia-all}. 

\clearpage

\clearpage

Likewise, $A$ and $B$ are each renormalized at one-loop order by the graphs illustrated in Fig.~\ref{Adiag} 
and Fig.~\ref{Bdiag} below, respectively.
\begin{widetext}

\begin{figure}[htb]
\includegraphics[width=14cm]{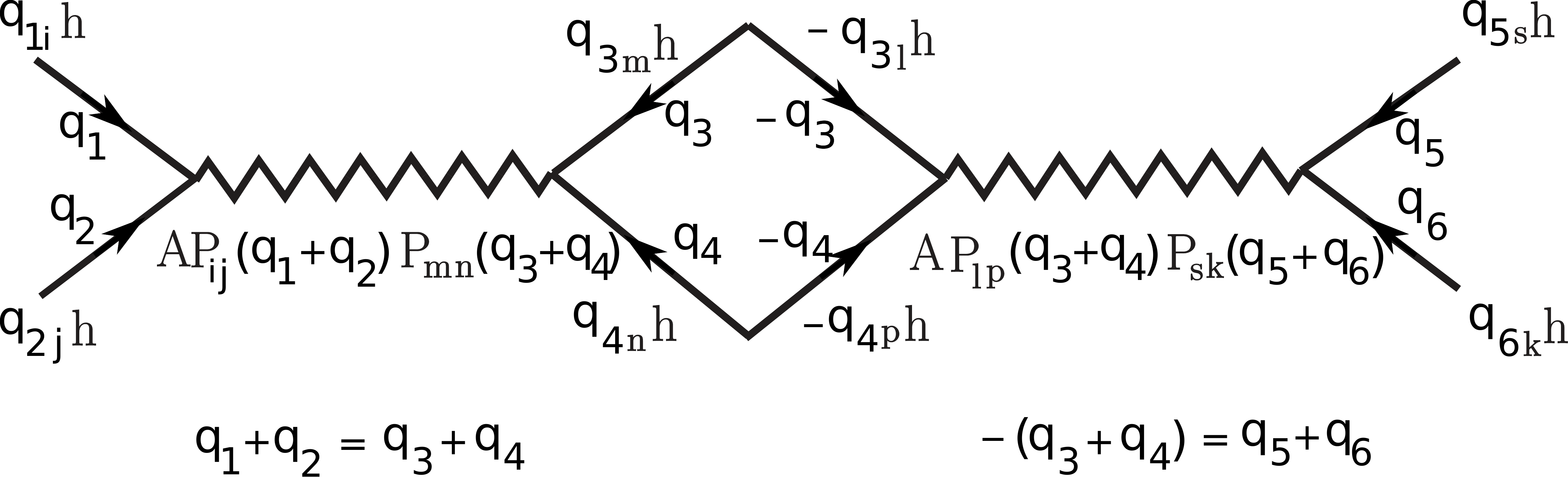}
\caption{One-loop Feynman diagram  for fluctuation corrections of 
$A$.}\label{Adiag}
\end{figure}

\begin{figure}[htb]
\includegraphics[width=14cm]{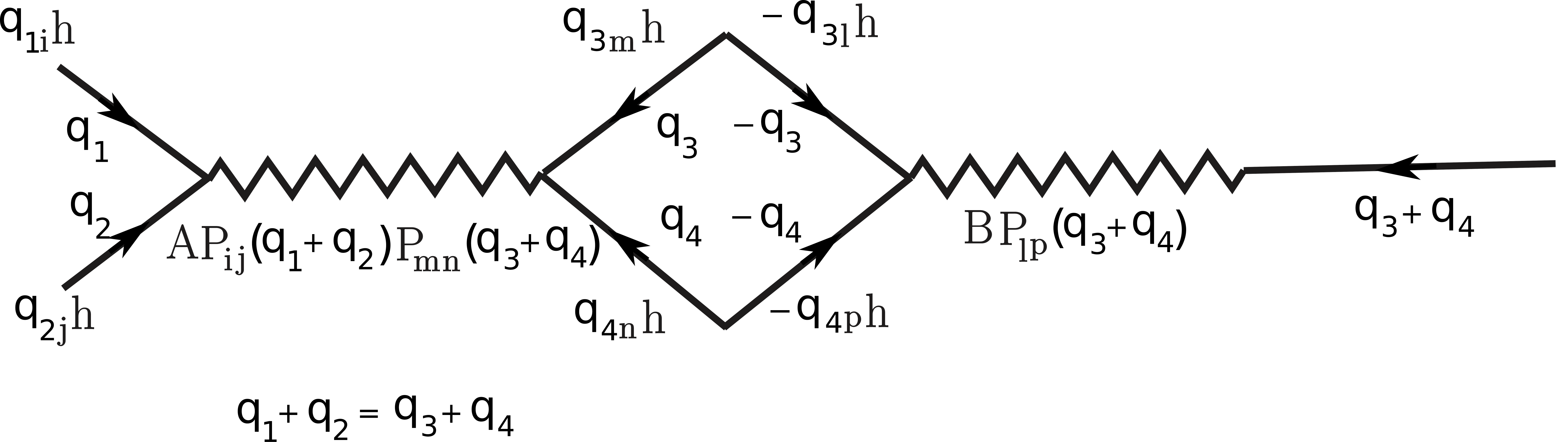}
\caption{One-loop Feynman diagram for fluctuation corrections of 
$B$.}\label{Bdiag}
\end{figure}

\end{widetext}


 
  Many one loop graphs that are topologically possible, e.g.,  Fig.~\ref{vanisha} and Fig.~\ref{vanishb}   in fact make vanishing contributions to $A$ and $B$, respectively. 
This is discussed in more detail in the Appendix, where we also calculate the graphs Figs.~\ref{kappa_dia-all}, \ref{Adiag}, and \ref{Bdiag} in detail.
The result is the following recursion relations:

\begin{equation}
 \frac{d\kappa}{dl}=\kappa\left[-\eta + g_1
   - \frac{5}{2}g_2\right]\,,
   \label{renor_kappa}
\end{equation}
\begin{equation}
 \frac{dA}{dl}=A\left[4-D-2\eta - \frac{g_1}{2}\right]\,,
   \label{renor_A}
\end{equation}
\begin{equation}
 \frac{dB}{dl}={B\over2}\left[4-D-3\eta - g_1\right]\,,   
 \label{renor_B}
\end{equation}
\begin{equation}
 \frac{dC_0}{dl}=(D-2+\eta)C_0 + g_2 \times {\cal O}(1)\,,\label{flowc}
\end{equation}
where we have defined two effective coupling 
constants, 
\beq
g_1\equiv\frac{AK_Dk_BT\Lambda^{-\epsilon}}{\kappa^2}\,\,\,,\,\,\,\,
g_2\equiv\frac{B^2K_Dk_BT\Lambda^{-\epsilon}}{ 
\kappa^3}\,\,,
\label{gdef}
\eeq
with 
$K_D=\frac{(D^2-1)S_D}{(2\pi)^DD(D+2)}$, where $S_D$ is the surface hyper-area of a D-dimensional sphere of unit radius, and $\epsilon\equiv4-D$. 
 We have not calculated the precise   value of the ${\cal O}(1)$ constant in (\ref{flowc}), as it 
affects none of the physics.

The recursion relations (\ref{renor_kappa}-\ref{flowc}) can be combined into a closed set of recursion relations for the dimensionless couplings 
 $g_1$ and $g_2$: 
  \beq
 \frac{dg_1}{dl}=g_1[\epsilon -\frac{5g_1}{2}+{ 5}g_2],\label{flow1}
 \eeq
 \beq
 \frac{dg_2}{dl}=g_2[\epsilon-4g_1+{{15\over2}}g_2]\,.\label{flow2}
\eeq

While we have derived these recursion relations to lowest order in $g_1$ and 
$g_2$, certain features of them are exact. These are: first, that the recursion 
relations for $g_1$ and $g_2$ 
are completely independent of the value of $C$. 
This is because $C$ does not enter the propagator, since it is a surface term,  
and is not a coefficient of a higher than  harmonic term.  Therefore, it does not affect 
the renormalization of the remaining model parameters. Second, the recursion 
relation (\ref{flowc}) for
$C$ becomes exact when $g_2\rightarrow 0$, because, once $g_2=0$ (which requires 
$\chi=0$), the Hamiltonian (except for $C$ itself) is completely 
inversion-symmetric, and, hence, contains no anharmonic terms that can generate 
an inversion-asymmetric term like $C$.

\begin{figure}[htb] 
\includegraphics[width=8cm]{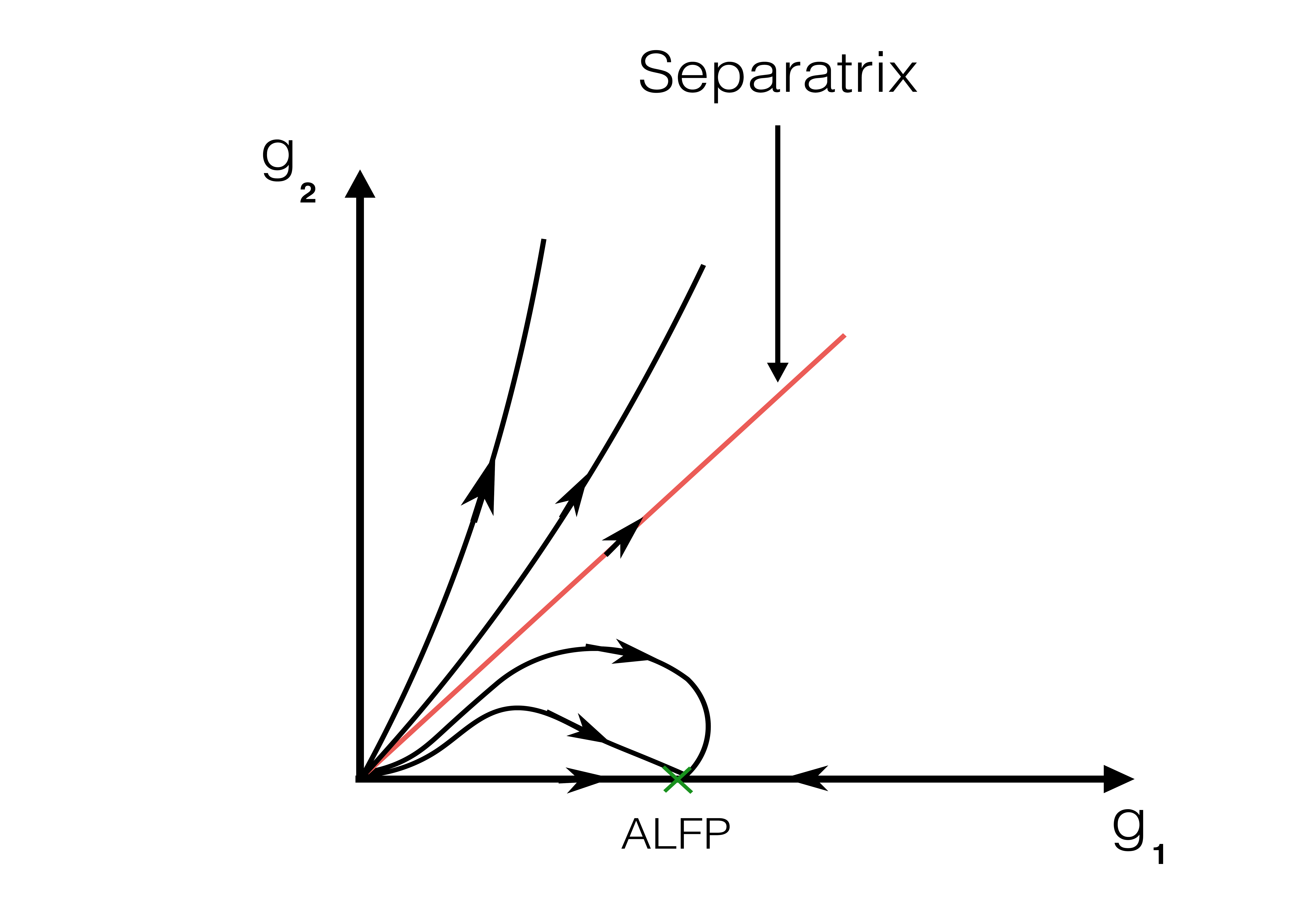}
 \caption{(Color online)  Schematic flow lines in the $g_1-g_2$ plane implied by the recursion relations (\ref{flow1}, \ref{flow2}). The 
 (green) cross marks the  only stable FP ($g_1=2\epsilon/5,g_2=0$). The   thick red curve is 
the 
separatrix  ${ g_2={3g_1\over5}}$ 
between the region controlled by the stable Aronovitz-Lubensky fixed point (``ALFP")~\cite{aronovitz1988}
(corresponding to   a spiral membrane) and the region  in which flows run off towards 
infinity.  We believe the membrane is crumpled in this region. }
\label{flow small}
 \end{figure}

 The RG flows implied by the recursion relations (\ref{flow1}, \ref{flow2}) are illustrated in Fig.~(\ref{flow small}). There are only two fixed points in the physical quadrant $g_{1,2}>0$: an unstable Gaussian fixed point at $g_1=g_2=0$, and a stable fixed point
\beq
g_1=2\epsilon/5,\,g_2=0 \,.
\label{stableFP}
\eeq

{ Since we derived our RG recursion relations  perturbatively assuming $g_1$ and $g_2$ were small, this result for the fixed point can only be trusted  if $\epsilon=4-D\ll1$. This is, of course, just the usual logic of the $\epsilon$-expansion \cite{wilson1975}. While we therefore do not expect our results to be quantitatively reliable all the way down to D=2, where $\epsilon=2$, we {\it do} expect the topology and general features of the flows to remain the same. In particular, we expect the long-wavelength physics of the membrane, for length scales large compared to all microscopic lengths, but much less than the length scale $L_H$ on which the membrane starts having self-avoidance interactions with itself,  to continue to be controlled by a fixed point at which $g_2=0$.
Because $g_2=0$ at this fixed point, asymmetry is {\em 
irrelevant} (in a 
scaling/RG sense) in the  phase that fixed point controls (i.e., for all systems whose ``bare" or initial $g_1(\ell=0)$, $g_2(\ell=0)$ lie in the basin of attraction of this stable fixed point).  Of course, this statement will cease to be true once the length scale under consideration grows to $L_H$, because at larger length scales the behavior of the membrane will be radically altered by self-avoidance interactions between successive turns of the spiral. But up to that length scale, asymmetry is irrelevant and the spontaneous curvature does not affect the physics of the membrane; therefore, the stable  fixed point (\ref{stableFP})  must be the same as the fixed point of a {\it symmetric} membrane; i.e., it must be the Aronovitz-Lubensky ~\cite{aronovitz1988} fixed point (which we will hereafter call the``ALFP") (hence the label ``ALFP" in Fig.~(\ref{flow small}))\footnote{The alert and well-informed reader will notice that both the position of this fixed point, and the value $\eta=2\epsilon/5$ of $\eta$ that keeps $\kappa$ fixed, are slightly different from those obtained by \cite{aronovitz1988}. This difference is simply due to the fact that we have analytically continued our model to dimensions D$>$2 in a slightly different way than  they did \cite{aronovitz1988}; our results should reduce to theirs in D=2,  where the ambiguity of continuation in dimension disappears}.

In particular, up to $L_H$, asymmetric membranes will  exhibit the same  anomalous elasticity of the bend modulus $\kappa$ as symmetric membranes. That is, the effective bend modulus $\kappa(\bq)$ at wavevector $\bq$ grows without bound as $\bq\to{\bf 0}$, diverging algebraically:
\beq
\kappa(\bq)=\kappa_0(q\xi_{_{NL}})^{-\eta} \,,
\label{kappaq}
\eeq
where $\kappa_0$ is the bare value of $\kappa$,  and $\xi_{_{NL}}$ is a non-universal length at which fluctuation corrections to $\kappa_0$ start to dominate over $\kappa_0$.  We will obtain  this length from our recursion relations below. Furthermore, $\eta$ is a universal exponent given by the value of our previously defined rescaling exponent $\eta$ required to keep $\kappa(\ell)$ fixed upon renormalization. At the ALFP near $D=4$, $\eta=g_1=2\epsilon/5$. We do not, of course, expect this result to be quantitatively accurate all the way down to the physical case D=2, for which $\epsilon=2$. However, it is reasonable to assume that asymmetry remains irrelevant all the way down to D=2 (again, this is the usual reasoning applied to any $\epsilon$-expansion, which assumes that the structure of the RG flows does not change as one moves to larger $\epsilon$). Hence, the value of $\eta$ that holds for asymmetric membranes will be the same as that for symmetric membranes in D=2; the best estimate of that value is provided by \cite{Leo}, which gives (\ref{eta}).

The analog of equation (\ref{kappaq}) in real space is the statement that the effective $\kappa$ on length scales $L$ obeys
\beq
\kappa(L)=\kappa_0\left({L\over\xi_{_{NL}}}\right)^{\eta} \,,
\label{kappaL}
\eeq
a result we will use later in our treatment of the spiral state.

We comment in passing that at this stable ALFP fixed point,  on length scales small enough that the membrane looks flat and isolated, the renormalized correlations of  the in-plane displacements $u_i ({\bf q})$ can be obtained  from (\ref{straindef}). We obtain {
\begin{equation}
 \langle u_i({\bf q})u_j(-{\bf q})\rangle=k_BT\left[\frac{P_{ij}({\bf q})}{\mu(q)q^2}+\frac{L_{ij}({\bf q})}
{(2\mu(q)+\lambda(q))q^2} \right] \,,
\end{equation}
where the longitudinal projection operator 
\beq
 L_{ij}(\bq)\equiv\frac{q_i q_j}{q^2}
 \label{Ldef}
 \eeq
projects 
any vector along  ${\bf q}$, and the 
renormalized elastic moduli $\mu(\bq)$ and $\lambda(\bq)$ are {\it vanishing} functions of wavevector $\bq$: 
\beq
\mu(\bq) \,, \lambda(\bq)\propto q^{\eta_\mu}
\label{muq}
\eeq
with 
\beq
\eta_\mu=4-D-2\eta \,.
\label{etamu}
\eeq
These results are  identical in every respect, including the value of $\eta_\mu$, to those for symmetric tethered membranes~\cite{aronovitz1988}.
}
}

The non-linear length $\xi_{_{NL}}$ is simply the length scale at which $\kappa$ starts to acquire appreciable fluctuation corrections. Equivalently, it is the length scale on which one or both of the dimensionless non-linear couplings $g_{1,2}$ become of $\cO(1)$. If the bare values $g^0_{1,2}$ of these couplings are both much less than $1$, then this length scale can be quite large. We will now estimate $\xi_{_{NL}}$ for the case in which the bare $g_{1,2}^0$ lie below the separatrix (\ref{strix}), and both are $\ll1$. 

Initially- that is, at renormalization group time $\ell=0$- the non-linear terms in the recursion relations (\ref{flow1}) and (\ref{flow2})   are negligible. Indeed, they will remain so at non-zero $\ell$ until the larger of $g_{1,2}(\ell)$ gets to be of $\cO(1)$. Thus, up to the value $\ell_1$ of $\ell$ at which this happens, the recursion relations in D=2 (where $\epsilon=4-D=2$) reduce to:
  \beq
 \frac{dg_1}{dl}=2g_1\,,\label{small g flow1}
 \eeq
 \beq
 \frac{dg_2}{dl}=2g_2\,,\label{small g flow2}
\eeq
whose solution is trivially
\beq
g_{1,2}(\ell)= g^0_{1,2}e^{2\ell} \,.
\label{glinsol}
\eeq

We can determine the value $\ell_1$ of $\ell$ at which the non-linearities become important by equating the larger of these to 1. This implies 
\beq
e^{\ell_1}={\rm min}\left({1\over\sqrt{g^0_{1,2}}}\right) \,.
\label{ell1.1}
\eeq
{ The non-linear length $\xi_{_{NL}}$ is just the length scale which, after precisely this much RG ``time", is rescaled to the inverse UV cutoff $\Lambda^{-1}$. This implies
\beq
\xi_{_{NL}}=\Lambda^{-1}e^{\ell_1}=\Lambda^{-1}{\rm min}\left({1\over\sqrt{g^0_{1,2}}}\right) \,.
\label{xinl}
\eeq
Using equation (\ref{gdef}) for $g_1$, with the parameters $\kappa$ and $A$ replaced by their bare values $\kappa_0$ and $A_0$ to obtain the bare value $g_1^0$ of $g_1$, and assuming that $g_1^0\gtrsim g_2^0$  (a condition which we'll show below applies throughout the spiral phase),  so that $g_1^0$ is the parameter that determines $\xi_{_{NL}}$, we obtain
\beq
\label{xiNLspiral}
\xi_{_{NL}}=\Lambda^{-1}e^{\ell_1}=\Lambda^{-1}\kappa_0\sqrt{1\over A_0 K_D \kbt}\Lambda^{\epsilon\over2}=\kappa_0\sqrt{1\over A_0  K_2 \kbt} \,,
\eeq
where in the last equality we have specialized to the physical case D=2, for which $\epsilon\equiv4-D=2$.}

The above discussion, and, in particular, equations (\ref{kappaq}) and (\ref{kappaL}), apply to all membranes whose bare parameters lie in the regime that flows upon renormalization into the symmetric ALFP fixed point. However, not all bare parameters do so. To see this, consider 
the evolution of the ratio $\frac{g_2}{g_1}$. The 
flow equations (\ref{flow1}) and (\ref{flow2}) imply 
\begin{equation}
 \frac{d}{dl}\ln \left(\frac{g_2}{g_1}\right)=\frac{1}{g_2}\frac{dg_2}{dl} - 
\frac{1}{g_1}\frac{dg_1}{dl}=\frac{{ 5}}{2}g_{ 2}-{3\over{ 2}}g_{ 1} \,.
\label{ratio}
\end{equation}
It is clear from this that if 
\beq
g_2={ {3\over5}}g_1 \,,
\label{strix}
\eeq
initially, this equality will continue to hold upon renormalization. Thus points on the locus (\ref{strix}) can {\it not} flow into the ALFP; indeed, they keep flowing out (to larger $g_{1,2}$) until they leave the regime of validity of our perturbation theory. Points above this locus can obviously not reach the ALFP either, since to do so, they would have to cross the locus (\ref{strix}), which they cannot do, since flow lines cannot cross. Therefore, the locus (\ref{strix}) acts as a separatrix between flows that go into the ALFP, which, as we have just discussed, imply scaling like symmetric membranes up to the length scale $L_H$, and those which instead flow out of the regime of validity of our perturbation theory, which will behave differently.

\begin{figure}
 \includegraphics[width=10cm]{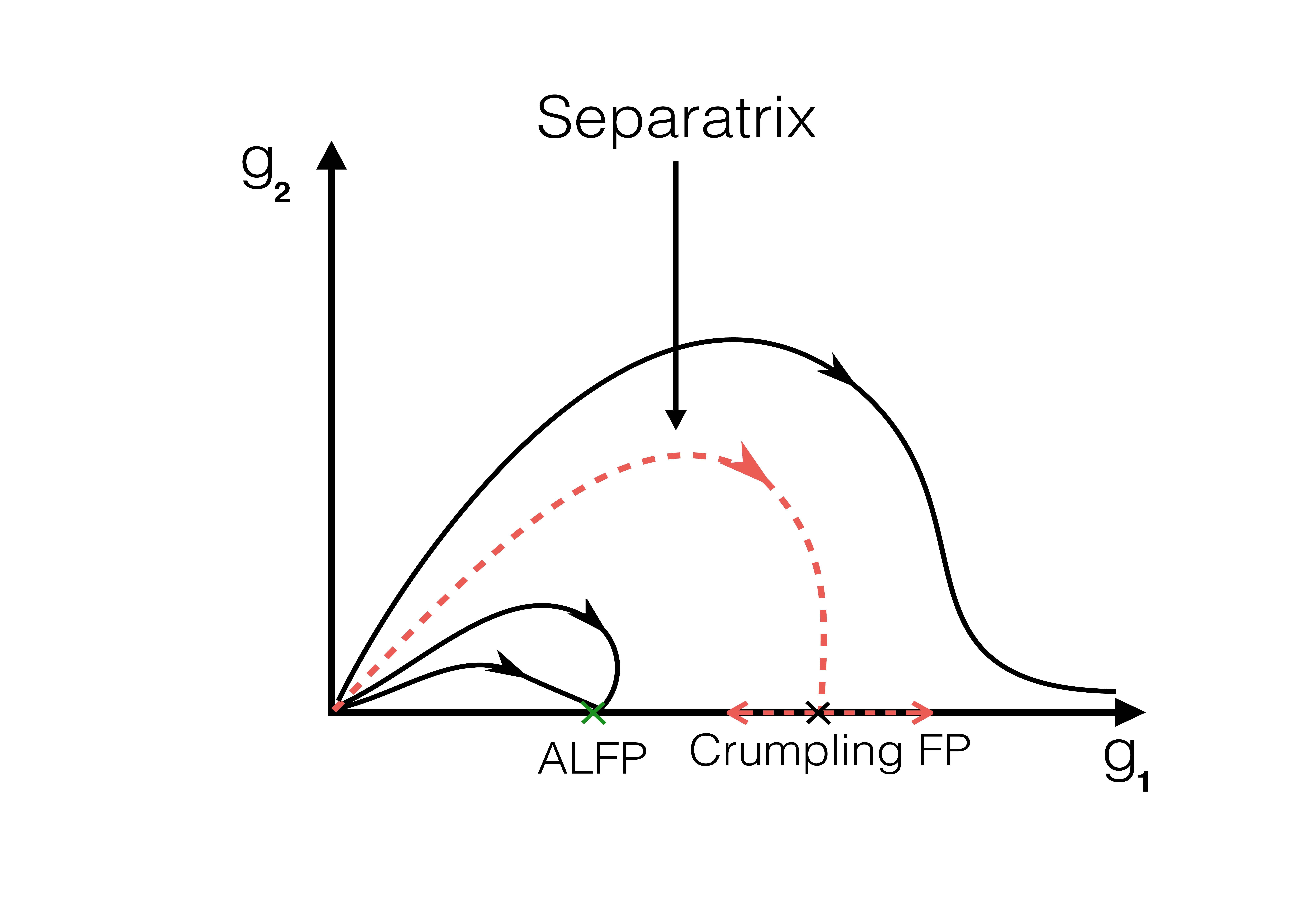}
 \caption{ Conjectured ``Occam's razor" global RG flows in the $g_1-g_2$ plane.}
 \label{fp}
\end{figure}

What is this different behavior? Since the flows in this regime lead out of the region of validity of our perturbation theory, we can only speculate. We will guide this speculation by the assumption that symmetric membranes have a continuous crumpling transition. This implies that at larger $g_1$ on the $g_1$ axis, there must be an unstable fixed point controlling this crumpling transition. If we now consider the full flows of an asymmetric membrane in the two dimensional parameter space $(g_1, g_2)$ and connect this putative flow on the $g_1$ axis with our flows near the origin in the simplest possible way (i.e., one that does not involve introducing any new fixed points), then we are lead to  Fig.~(\ref{fp}). This is an ``Occam's razor" argument:  Fig.~(\ref{fp}) is the simplest flow topology that incorporates our known flows for small $g_{1,2}$ and $\epsilon$ with the putative flows that allow for a continuous  crumpling transition for a symmetric membrane.
Note that this conjectured topology implies the separatrix flows to the crumpling fixed point; this implies that if we vary membrane parameters (including temperature) in such a way that the bare $g_{1,2}$ cross the separatrix, the membrane will crumple. 

This interpretation of the separatrix as the crumpling threshold is supported by the observation
that the underlying reason for the runaway RG flows on and above the separatrix is that the bend modulus $\kappa$ is being driven downwards by the $-{5\over2}g_2$ term in its recursion relation. If $\kappa$ is driven to zero by this term, $g_1$ and $g_2$ both diverge. If  $\kappa$ is driven to {\it negative} values by this term, the membrane will clearly crumple. 

We cannot, of course, follow this flow of $\kappa$ from positive to negative values, since the divergence of $g_1$ and $g_2$ as $\kappa\to0$ invalidates our perturbative RG calculation. 
But the structure of the recursion relations just discussed strongly suggests that the separatrix is the crumpling boundary, which supports our conjectured topology Fig.~(\ref{fp}) for the global RG flows.

Note that this crumpling is occurring in a regime in which the bare bend modulus $\kappa_0$ given by eqn. (\ref{kappa})  is positive; this is an anharmonic crumpling mechanism, beyond the harmonic theory of crumpling we developed in section (\ref{QT}). Note also $\kappa (\ell)>0$ at small enough $\ell$; this implies small enough membranes will not crumple, which in turn implies our two length scale regimes in fig. (\ref{k-chi}). 
We will discuss this in more detail in section (\ref{Crumpling revisited}).

\section{$L\gg L_H$: Global membrane structure, the spiral state, and crumpling revisited}\label{L gt Lfree} 

\subsection{The spiral state for $T=0$}

Since the $A$ term is long-ranged, and a perfect square, 
this  implies that, in the absence of the $B$ term, the lowest energy 
configurations of the membrane will have zero 
Gaussian curvature. This is why the "spiral" state of the membrane, that 
results   from the  competition between the spontaneous curvature, bending energy and 
excluded volume effects, curls in 
only 
one direction. We now heuristically argue that at $T=0$, the $B$-term, 
even when 
nonzero,  is always irrelevant in  the thermodynamic limit, making the 
spiral phase  the only ground state at $T=0$. To show this,  we   make a rough estimate of the free energy  
${\mathcal F}_h$ in (\ref{free energy_h_RS}) in terms of  
Gaussian curvature $S$ and  mean curvature $M$, both of which are assumed to
be constant for simplicity.  this gives
\begin{equation}
 {\mathcal F}_h^{\rm rough}= { \kappa_0}L^2(M-\frac{C}{\kappa_ 0})^2 + A_0L^6 S^2 
+ B_0 L^4MS\,,\label{fscheme}
\end{equation}
where we have  ignored numerical factors of $\cO(1)$, and logarithmic factors, in this rough estimate, and have   crudely estimated the kernels of the $A_0$ and $0$-terms in 
${\mathcal F}_h$ (\ref{free energy_h_RS})  as $L^2$ and $L^0$ respectively. Minimizing 
${\mathcal F}_h^{\rm rough}$ with respect to  the Gaussian and mean curvatures $S$ and $M$ yields, again ignoring factors of $\cO(1)$,
\begin{eqnarray}
 { -\kappa_0} L^2 (M-\frac{C}{\kappa_0})&=&B_0L^4S \,,\\
-A_0 L^6 S &=&B_0 L^4M \,.
\end{eqnarray}
These are readily solved to give 
\begin{eqnarray}
 M&=&{C\over \kappa_0-{B_0^2\over A_0}} \,,\\
 S &=& -{B_0C\over A_0\kappa_0-B_ 0^2}{1\over L^2} \,.
\end{eqnarray}
Thus in the thermodynamic limit $L\rightarrow \infty$, $S$ vanishes and the 
ground state must be a state given by $M=C/\kappa_0\equiv2r_0$  with zero Gaussian 
curvature, where $r_0$ is given by (\ref{r011}); see also Appendix~\ref{hole1}. 
 This implies that the membrane must bend only in one direction; the inversion symmetry breaking $B$ term does not alter this conclusion.
 {
For a square membrane, the bent direction 
is {\em spontaneously} chosen; for a rectangular membrane it is energetically 
profitable to roll up along the longer direction.

The
above argument assumes that the bend modulus $\kappa_0$ is positive. As we have seen, there are regions of the parameter space for which this is not true: specifically, $\chi^2>\chi_U^2$. In that case, the membrane wants to maximize both its Gaussian and its mean  curvatures, which it does by crumpling.

Returning now to the case  $\kappa_0>0$, we note that, }
while it will not affect the conclusion that the membrane will bend in only one direction, self-avoidance will radically alter the radius of curvature in the single bent direction.
{ This becomes obvious once we note that, were the membrane to roll up into a cylinder with radius $r_0$, its volume $V$ would be  $V\sim r_0^2 L <<L^2 a$, where $L^2a$ is the volume of the material of the membrane itself.  Therefore, such a tightly rolled membrane would be extremely self-overlapping. To avoid this, it must wrap less tightly. On the other hand, it energetically prefers to be wrapped as close the optimal radius of curvature $r_0$ as possible. It can do this by wrapping as tightly as possible in one direction without overlapping. The structure that results can be seen by 
imagining starting with flat membrane, and wrapping it  from one end at spontaneous curvature. When one has rolled up a length $2\pi r_0$ of membrane, the end of the membrane encounters 
the remainder (i.e., the as yet unrolled up portion) of the membrane.
This part therefore cannot wrap at $r_0$, so it instead wraps up as tightly as it can, which is readily seen to be a radius of curvature $r_0$ plus $a$. This continues until this section is rolled up into the remainder of membrane; now radius of curvature becomes $r_0+2a$. 
Each successive turn is therefore spaced by the membrane thickness $a$ from the previous one, leaving just enough room for one layer. This is clearly the tightest wrapping allowed by self-avoidance. This structure we've just described  is a spiral of Archimedes (\ref{rspiral}), with a hole in the center of radius $r_0$. In appendix VI, we show that the radius of this hole is indeed $\cO(r_0)$; in fact, it is   $r_0$, 
which is therefore negligible for $L>>r_0$.
F

}
 
 \subsection{The spiral state for $T\ne0$}

We now turn to the effects of thermal fluctuations on the spiral 
phase. This requires studying the system at larger scales $L\gg 
L_H$. That a large enough asymmetric membrane takes the form of a 
double spiral (see Fig.~\ref{dspiral}), should still hold for $T>0$. 
 Thermal fluctuations, however, considerably affect the form of the spiral
 by giving rise to a longer ranged "Helfrich repulsion"\cite{helfrich1978} that 
has its origin in excluded volume interactions and important over scales $L\gg 
L_H$. This
opens the spiral up. When such fluctuations are important (as they always will 
be for a sufficiently large membrane), the form of the spiral changes to a 
power 
law, as we show below.
 {The Helfrich interaction energy at $T>0$ was first derived for fluid 
membranes in \cite{helfrich1978}, and was 
calculated for tethered membranes in~\cite{toner1990}. 

We review this calculation here. Up to the length scale $L_H$ at which the membrane starts interacting with neighboring turns of the spiral, it acts like a free membrane, as treated in the last subsection. Therefore, the contribution of fluctuations on shorter length scales to the total mean squared height fluctuations  $\left<h(\br)^2\right>$
can be calculated precisely as one would for a free membrane, but with an infrared cutoff (i.e., minimum wavenumber) given by the inverse of $L_H$. This implies:
\bea
\left<h(\br)^2\right>=\int_{q>q_m}{d^2q\over(2\pi)^2}\,\left<|h(\bq)|^2\right> \,,
\label{h(r) flucs}
\eea
with the infrared cutoff $q_m\sim1/L_H$.
Our RG analysis showed that, at small $\bq$, which is readily seen to be the regime of wavevector that dominates the integral in this expression (\ref{h(r) flucs}),
\beq
\left<|h(\bq)|^2\right>={\kbt\over\kappa(\bq)q^4} \,,
\label{hq}
\eeq
with the effective, renormalized bend modulus $\kappa(\bq)$ given
by (\ref{kappaq}), 
with, we recall, the anomalous elasticity exponent $\eta$ precisely the same as that for {\it symmetric} membranes, in the regime in which our RG flows go into the symmetric ALFP in figure (\ref{fp}). Since
the integral over $\bq$ in (\ref{h(r) flucs}) is dominated by small wavenumbers $q$, we obtain
\beq
\left<h(\br)^2\right>={\kbt\over\kappa_0}\xi_{NL}^\eta L^{2-\eta}_H\times \cO(1) \,,
\label{h(r) flucs2}
\eeq
where we have absorbed our uncertainty about the precise value of the infra-red cutoff into the $\cO(1)$ factor.
We can now obtain $\lcn$ by roughly equating this mean  squared fluctuation to the square of the distance $d$ to the  next turn, since a patch of membrane of this size is just big enough to fluctuate enough to contact the turn above it. This gives
\beq
\left<h(\br)^2\right>={\kbt\over\kappa_0}\xi_{NL}^\eta L^{2-\eta}_H\times \cO(1)= d^2\times \cO(1) \,,
\label{Lc cond}
\eeq
which is trivially solved for $\lcn$:
\beq
\lcn=\left({\kappa_0\over\kbt}\right)^{1/(2-\eta)}\xi_{NL}^{-\eta/(2-\eta)} d^{2/(2-\eta)}\times \cO(1) \,.
\label{lcn}
\eeq

With this result for the typical distance $\lcn$ between points of contact between neighboring membranes in hand, one can now argue~\cite{helfrich1978} that each such contact causes a reduction in entropy, since the motion of the membrane is restricted by self-avoidance at these points. Assuming this reduction is of $\cO(1)$ for each contact implies that each contact costs a typical free energy of $\cO(\kbt)$. Thus the total free energy cost per unit area is
given by
\bea
\mathcal{U}_H( d)={\kbt\over L^2_H}\times\cO(1)=\left({k_BT \over \kappa_0}\right)^2A_0 \left({w \over 
 d}\right)^\gamma\times\cO(1), \nonumber\\\label{hel}
\eea

where 
\beq
\gamma\equiv {4 \over 2-\eta} \,,
\label{gammadef}
\eeq
and we have used our earlier expression (\ref{xiNLspiral}) for $\xi_{NL}$ to write this expression entirely in terms of 
the bare values $\kappa_0$ and $A_0$ of $\kappa$ 
and $A={4\mu(\mu + \lambda) \over 2\mu + \lambda}$, and the length $w$ is 
given by 
\beq
w=\sqrt{\kappa_0 \over A_0} \,.
\label{wdef}
\eeq
In a simple model of the membrane as an elastic continuum one would obtain~\cite{Landau}  $A_0\sim\mu_{3d}a$ and $\kappa_0\sim\mu_{3d}a^3$, which imply $w=a\times\cO(1)$, where $a$ is the membrane thickness.

 Now let us consider the effects of this interaction on   a spiral membrane.  We first need to relate the distance $d$ between successive turns of the spiral to its radius profile $r(s)$, where $s$ is arc-length along the spiral from the center. Note that in general, unlike the Archimidean spiral, this distance will vary with arc-length $s$ along the spiral.
 
 If the spiral   is very tightly wound,  (and we will verify {\it a posteriori} that it is), so that the angle between the spiral and the radius drawn from the 
center of the spiral to the point in question is close to $90\degree$, then the spacing $d(s)$ of successive turns of the membrane  at a distance $s$ along the membrane 
is the difference between $r(s=s_A)$ and $r(s=s_B)$,  as illustrated in Fig.~\ref{spiral2}. 

\begin{figure}[htb]  
 \includegraphics[width=6cm]{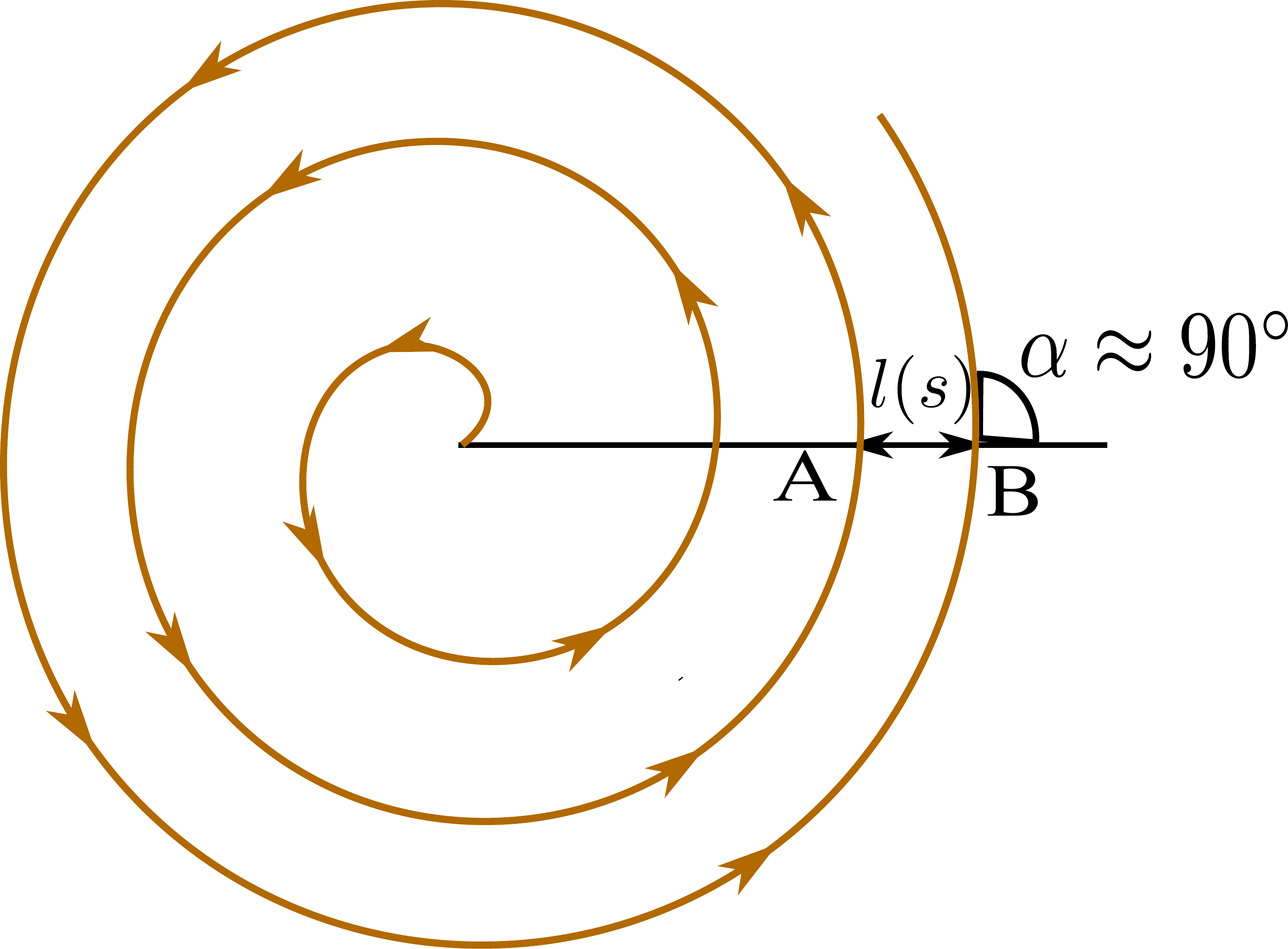}
 \caption{(Color online) Cross section of a single spiral at $T>0$.}  
\label{spiral2}
 \end{figure}
Note that $s_B-s_A$ is very nearly $2\pi r(s)$ for a tightly wound spiral (i.e., 
the path between $A$ and $B$ is nearly a circle of radius $r(s_A)$). Furthermore,  
if ${dr  \over ds}$ varies slowly with $s$  (which we will again verify {\it a posteriori}), so that it is nearly constant 
between $A$ and $B$, then we can say that the spacing between successive turns of the membrane is given by
\bea
d(s_A)\approx r(s_B)-r(s_A)\approx 2\pi r(s_A){dr \over ds}|_{s=s_A}\,, \nonumber\\
\label{ls}
\eea
or more generally
\bea
d(s)\equiv 2\pi r(s){dr \over ds}\,. \label{lgen}
\eea
 The structure of the spiral can now be determined by balancing this Helfrich interaction against the  two other terms in the free energy: the  spontaneous curvature energy $C/R$  (where $R$ is the mean curvature), and the bending energy. The latter must be calculated using the renormalized value of $\kappa$. In doing so, we must recognize that the anomalous  length  dependence (\ref{kappaL}) of $\kappa$  is cut off for  length scales  $L\gtrsim L_H$, since the height fluctuations stop growing at that point, due to the self-avoidance interactions with the next turn of the spiral. This implies that $\kappa$ becomes length scale independent larger length scales. 
Matching this constant onto the value of $\kappa(L)$ at the  largest $L$'s for which the elasticity is still anomalous, namely $L\sim L_H$, implies that this constant value of $\kappa$ on larger length scales is given by
\beq
\kappa(L>L_H)\sim\kappa(L\sim L_H)\sim \kappa_0 \left({L_H \over \xi_{_{NL}}} \right)^\eta\,.
\label{kappaLH}
\eeq
Since  $L_H$ increases with $d(s)$, where $d$ is the distance to the next turn of the membrane, the fact that $d(s)$ depends on distance $s$ along the spiral implies that $L_H(s)$ will as well. Hence, through, (\ref{kappaLH}), so will $\kappa$.

The spontaneous curvature coefficient $C$, on the other hand, exhibits no such anomaly, since, as we noted in our RG discussion in the last section, its graphical corrections vanish when $g_2$ does, as it does in the region of parameter space we are considering here.

The final ingredient we need to calculate the bend and spontaneous curvature energies is an expression for the mean curvature $R(s)$, which will also depend on distance $s$ along the spiral. For a very tightly wound spiral, this is simply $r(s)$ itself. 

We can summarize all of the above reasoning in the following expression for the energy of a spiral:
\bea
E=2L_{||}\int^{L_\bot/2}_0 ds \left[-{C \over r(s)} + {\kappa_0 \over 
{ 2}r^2}\left({L_H(s) \over \xi_{_{NL}}} \right)^\eta + {\mathfrak{D} \over [r{dr \over ds}]^\gamma} 
\right], \nonumber\\
\label{energyhel}
\eea 
where 
\beq
\mathfrak{D}=\left({k_BT \over \kappa_0}\right)^2{A_0 w^\gamma \times\cO(1)} \,.
\label{Ddef}
\eeq
 
We will now obtain the structure $r(s)$ of the spiral by minimizing this free energy. In doing so, we will assume, and verify {\it a posteriori}, that the  bending energy is negligible for a sufficiently large spiral. Doing so, 
the Euler-Lagrange equation for $r(s)$ 
obtained by minimizing the energy (\ref{energyhel}) over $r(s)$ is
\bea
-\gamma \mathfrak{D}{d \over ds}\left[{1 \over r^\gamma ({dr \over ds})^{\gamma +1}} 
\right]= {C \over r^2} - {\gamma \mathfrak{D} \over r^{\gamma+1}({dr \over ds})^\gamma} \, . 
\label{el}
\eea
Formidable though this equation looks, it is easily solved by the ansatz: 
\bea
r=s_0\left({s \over s_0}\right)^\alpha\,, \label{ralpha2} 
\eea
which, when inserted into the Euler-LaGrange equation (\ref{el}) yields

\begin{widetext}
\beq
\Gamma_1(\alpha,\gamma) \mathfrak{D} s_0^{[(\alpha-1)(2\gamma+1)]}s^{-[\alpha(2\gamma+1)-\gamma]}=Cs_0^{2 (\alpha-1)}s^{-2\alpha}
\label{elans1}
\eeq
\end{widetext}
where we have defined the unimportant, $\cO(1)$ constant 
\beq
\Gamma_1(\alpha,\gamma)=\gamma \alpha^{-(\gamma+1)}[2(\gamma+1)\alpha-\gamma-1]\,.
\label{Gamma_1def}
\eeq


{ Balancing powers on both sides of this equation determines the shape exponent $\alpha$:
\bea
\alpha={\gamma \over 2\gamma-1}\,, \label{alphascl}
\eea
while equating the constant prefactors determines $s_0$:

\beq
s_0=\left(\mathfrak{D}\over C\right)^{1\over\gamma-1}\Gamma_2(\gamma)\,,
\label{s0}
\eeq
where we have defined yet another unimportant $\cO(1)$ constant 
\beq
\Gamma_2(\gamma)=\left[(\gamma+1)\left({2\gamma-1\over\gamma}\right)^\gamma\right]^{1\over\gamma-1}\,,
\label{Gamma_2def}
\eeq
which is a monotonically decreasing function of $\gamma$ over the entire allowed range $2<\gamma<4$, and hence bounded above by $\Gamma_2(2)=6.75$ exactly, and below by $\Gamma_2(4)\approx3.606125$. If we take the canonical value of $\eta={4\over1+\sqrt{15}}$, which implies $\gamma\approx3.39$, we get $\Gamma_2\approx3.95655$. 
}

Using our expression (\ref{gammadef}) for $\gamma$ in terms of $\eta$ 
 in (\ref{alphascl}) gives
\bea
\alpha={4 \over 6+\eta}. \label{alpha}
\eea

 Using these expressions (\ref{s0}) and (\ref{alpha}) in our ansatz (\ref{ralpha2}) for $r(s)$ completely specifies the shape of the spiral. To express that shape in more familiar polar coordinates,  we use the fact that, for a tightly wound spiral, 
\bea
{ds \over d\theta}=\sqrt{r^2+\left({dr\over d\theta}\right)^2}\approx r \,. \label{arclength0}
\eea

Using  Eq.~(\ref{ralpha2}), this
can be integrated to give

\bea
\left({s\over s_0}\right)^{1-\alpha}= (1-\alpha) \theta, \label{seq}
\eea
with the boundary condition $\theta=0,\,s=0$.
Thus solving for $s(\theta)$: 

\bea
s  = s_0((1-\alpha)\theta)^{1/(1-\alpha)} \,. \label{seqfin}
\eea

Using Eq. (\ref{seqfin}) in (\ref{ralpha2}), 
we get $r(\theta)$ as a function of $\theta$:
\bea
r(\theta)=R_0\theta^\nu, \label{rtheta1} 
\eea 
where 
\beq
\nu={\alpha \over (1-\alpha)}={4 \over 2+\eta}{ 
\approx4-{2\over3}\sqrt{15}\approx1.418} \,,
\label{nu2}
\eeq
and 
\beq
R_0\equiv s_0(1-\alpha)^\nu \,.
\label{R0}
\eeq
The numerical estimates of the exponents are 
based on the theoretical estimate of $\eta$   for the flat phase of 
asymmetric membranes:
\bea
\eta={4\over 1+\sqrt{15}}\approx .821
\eea
 obtained by Radzihovsky and LeDoussal\cite{Leo}. 
 
  The above results are those summarized in equations (\ref{rtheta}), (\ref{nu}), and (\ref{eta}) of the introduction.

Using our earlier result  (\ref{s0}) for $s_0$ in (\ref{R0}), we find that the scale length $R_0$ exhibits universal scaling with 
temperature and other parameters, which can also be related exactly to the 
exponent $\eta$; we find
\bea
R_0=\left[(k_B T)^{2(2-\eta)}\kappa_0^{2(\eta-1)}A_0^{-{\eta}}C^{(\eta-2)}\right]^{1/(2+\eta)}\times\cO(1) \,,\nonumber\\
\label{R_01}
\eea
where  we remind the reader that  $\kappa_0=\kappa(l=0)$ is  the "bare" bend modulus  and 
$A_0\equiv\frac{4\mu_0(\mu_0+\lambda_0)}{2\mu_0+\lambda_0}>0$, with $\mu_0$ and 
$\lambda_0$  the equally bare Lame' coefficients. Here, by ``bare", we mean the values these parameters have 
before being renormalized by thermal fluctuation effects.
 In deriving (\ref{R_01}), we
 have used our expressions (\ref{Ddef}) for $D$ and (\ref{wdef}) for $w$.
 This result (\ref{R_01}) is precisely equation (\ref{R_0}) of the introduction.

The total radius $R_T$ of the spiral regions also exhibits universal scaling, in 
this case  with
the spatial extent $L$ of the membrane, as may be obtained from (\ref{ralpha2}) with $s=L$ 
\bea
R_T=R_0^{1-\alpha} L ^\alpha \,.
\label{ralpha1} 
\eea


It is straightforward
to use these results to verify our three earlier {\it a posteriori} assumptions, which we remind the reader were:

\noindent 1) that the spiral is ``tightly wound", in the sense that the angle between the spiral and the radius vector is close to $90$ degrees,  

\noindent 2) that ${dr\over ds}$ does not vary appreciably between successive turns of the spiral, and

\noindent 3) that the bending energy is negligible compared to the Helfrich interaction and the spontaneous curvature energy.

Both of these results follow from the algebraic (i.e., power-law) form 
of the spiral. The angle $\phi$ between the radius vector and the spiral obeys
\beq
\tan\phi={d\theta\over d\ln r} \,.
\label{phi}
\eeq
Using our expression (\ref{ralpha2}) for $r(s)$ in this expression, we see that 
\beq
\tan\phi={\theta\over \nu} \,,
\label{phi2}
\eeq
which is clearly $\gg1$ after the first few turns (i.e., for $\theta\gg2\pi$), since $\nu\approx1.418$ is $\cO(1)$. This implies that 
\beq
\phi\approx{\pi\over2}-{\nu\over\theta} \to{\pi\over2}\,.
\label{phi3}
\eeq

This completes our demonstration that the spiral is tightly wound. Turning next to the question of whether ${dr\over ds}$ varies appreciably over one turn of the spiral, we note that we can estimate the change in ${dr\over ds}$ over one turn as
\beq
\Delta\left({dr\over ds}\right)=\left({dr\over ds}\right)_{s+2\pi r}-\left({dr\over ds}\right)_{s}\approx 2\pi r {d^2r\over ds^2} \,.
\label{rdot}
\eeq
Using (\ref{ralpha2}) in this expression yields 
\beq
\Delta\left({dr\over ds}\right)= 2\pi \alpha(\alpha-1)\left({s\over s_0}\right)^{2(\alpha-1)}\to 0 \,,
\label{rdot2}
\eeq
as $s\to\infty$, since $\alpha<1$. 

Finally we turn to our third {\it a posteriori} assumption, that  the bending energy is negligible compared to the Helfrich interaction and the spontaneous curvature energy.

The bending free energy density, as shown in (\ref{energyhel}), is
\beq
f_{\rm{bend}}={\kappa_0 \over 
r^2(s)}
\left({L_H(s) \over \xi_{_{NL}}}\right)^\eta \,.
\label{fb1}
\eeq

Using our expression (\ref{lcn}) for $L_H$ in terms of the spacing 
$d$ between successive turns of the membrane, and using (\ref{lgen}) to relate $d(s)$ to $r(s)$,  we find that
\beq
f_{\rm{bend}}=\Gamma_B r^{4(\eta-1)\over2-\eta}\left({dr\over ds}\right)^{2\eta\over2-\eta} \,,
\label{fb2}
\eeq
where we have defined the constant
\beq
\Gamma_B \equiv\left({\kappa_0\over \kbt}\right)^{2\over2-\eta}\xi_{_{NL}}^{-\left({2\eta\over2-\eta}\right)}\kbt \,.
\label{Gbdef}
\eeq

Now using our expressions (\ref{ralpha2}) and (\ref{alpha}) for $r(s)$ in (\ref{fb2}), we find that 
\beq
f_{\rm{bend}}\propto s^{\nu_b} \,,
\label{fb3}
\eeq
where the exponent $\nu_b$ is  given by
\beq
\nu_b={2(\eta-4)\over6+\eta} \,.
\label{nub}
\eeq

Likewise, the spontaneous curvature free energy density $f_C\equiv -C/r(s)$ obeys
\beq
f_C\propto {1\over r(s)}\propto s^{-\alpha}=s^{-\left({4\over6+\eta}\right)} \,.
\label{fc}
\eeq
Taking the ratio of the bending energy (\ref{fb3}) to this, we find
\beq
{f_{\rm{bend}}\over f_C}\propto s^{\nu_{bc}} \,,
\label{bc}
\eeq
where
\beq
\nu_{bc}={2(\eta-4)\over6+\eta}+{4\over6+\eta}={2(\eta-2)\over6+\eta} \,.
\label{nubc}
\eeq
Since $\eta<2$, this exponent $\nu_{bc}<0$. Therefore, as $s\to\infty$, the ratio of the bending energy to the spontaneous curvature energy vanishes, proving that it is, as we assumed, negligible for a sufficiently large membrane.

This completes our {\it a posteriori} verification of all three of the assumptions we used in deriving the form of the spiral.

We now argue that asymmetric tethered membranes in the spiral phase indeed display long range order. By long range order we mean {\em predictability} of the direction of the local normal $ {\bf n}$  throughout the membrane, given its position at one point on the membrane, in the thermodynamic limit. This requires that the variance of the fluctuations $\delta \bf n$ of the local normal about its position at $T=0$ must be bounded in the thermodynamic limit. This is true for statistically flat symmetric tethered membranes, for which the local normals are all parallel to each other at $T=0$. For asymmetric tethered membranes at $T=0$, the normals are not parallel  due to the spiral structure; nonetheless, they are uniquely   {\it determined} by that spiral structure everywhere on the membrane. We now argue that the variance $\langle (\delta {\bf n})^2\rangle$  about this deterministic spiral structure is indeed finite in the spiral phase of asymmetric tethered membranes. 

We  begin by noting that $\langle(\delta {\bf n})^2\rangle$ gets contributions from two regimes of wavevectors: 

\noindent(i) 
$ q\gg L_H^{-1}$, on which the elasticity of the membrane looks like that of a symmetric membrane in isolation. Therefore, the contribution of fluctuations from this range of wavevector to $\langle(\delta {\bf n})^2\rangle$ is finite  for the same reason  - namely, the divergence of $\kappa(\bq)$ at long wavelengths- as for symmetric tethered membranes.

\noindent(ii) $q\ll L_H^{-1}$, on which the elasticity of the membrane looks like that of a bulk smectic A. For that range, the standard theory of smectic layer fluctuations implies\cite{chaikin2000} that
\beq
\langle (\delta {\bf n})^2\rangle=\int {d^3q\over(2\pi)^3} \,\,{k_BT q_\perp^2\over(\tilde B q_z^2+Kq_\perp^4)} \,.
\label{smecdirfluc}
\eeq}
Here, $\tilde B$ and $K$ are  respectively the standard layer compression and layer bending elastic constants for smectics, which are related to $\kappa (L_H)$: $K\sim \kappa (L_H)/d$ and $\tilde B = {\partial^2 U_H\over\partial d^2}d\sim  ( \kbt)^2/(\kappa (L_H) d^3)$;  and $q_z$  and $q_\perp$ are respectively the magnitudes of the  components of the wavevector along  and perpendicular to the direction locally perpendicular to the layers. 
The integral in (\ref{smecdirfluc}) converges down to ${\bf q}=0$. Thus $\langle (\delta {\bf n})^2\rangle$ remains finite; this establishes that  long range orientational order exists in the spiral phase. 

 The above analysis applies on length scales small compared to the local radius of the spiral. On longer length scales, the director simply follows the normal to the spiral.


\subsection{Crumpling revisited}
\label{Crumpling revisited}

 Having discussed the spiral phase, 
we turn now  to the other
region of parameter space, namely that which flows away 
from the ALFP, and towards negative $\kappa$. 
While 
we cannot follow these flows all the way to $\kappa
=0$ (since both $g_{1,2}$ diverge there, so that our perturbation theory breaks down), we suspect that this signals crumpling of 
large membranes. This region of parameter space  therefore corresponds to the 
crumpled phase. 
For $\epsilon=4-d\ll1$, which is the region in which our perturbative RG is 
accurate, this is the region in Fig.~(\ref{k-chi}) lying
above the separatrix   $g_2=3g_1/5$. 
For the physical case $\epsilon=2$, it seems 
reasonable to assume that there continues to be a separatrix which, for small 
$g_{1,2}$, is a straight line 
$g_2=\rho g_1$
of universal slope  $\rho=\cO(1)$, although since $\epsilon=2$ we  cannot
calculate the universal constant $\rho$. 

The range of bare asymmetry parameter 
$\chi$ in our original model (\ref{free energy}) that we are now discussing is 
$\chi_{_{_L}}^2<\chi^2<\chi_{_U}^2$, where the upper bound follows because we are 
considering positive $\kappa _0$ in equation (\ref{kappa}), while the lower bound follows 
from assuming that we are above the separatrix, which implies, for small bare 
$g^0_{1,2}$, that $g^0_2/g^0_1>\rho $. Using our earlier expressions (\ref{gdef}) for 
$g_{1,2}$, we see that this implies
\beq
\chi^2>{\rho\kappa'(2\mu_0+\lambda_0)\over\rho+{\mu_0\over\mu_0+\lambda_0}}
={\chi_{_U}^2\over 1+{\mu_0\over\rho(\mu_0+\lambda_0)}}\equiv\chi_{_{_L}}^2 \,,
\label{chiL3}
\eeq
where in the equality we have used our result  (\ref{chiu}) for $\chi_{_U}^2$.
Note that, reassuringly, we always have $\chi_{_{_L}}^2<\chi_{_U}^2$, since $\rho$, $\mu_0$, and $\mu_0+\lambda_0$ are all positive, the latter two positivities being required for stability.

For $\chi$'s in the range $\chi_{_{_L}}^2<\chi^2<\chi_{_U}^2$, the membrane can remain  uncrumpled if it is sufficiently small. 
This is because the 
bare value $\kappa(\ell=0)=\kappa_0=\kappa'-\frac{\chi_0^2}{2\mu_0+\lambda_0}$
 is positive  in this  range of $\chi$'s, and can stabilize orientational order, and thereby prevent crumpling,  on length scales short enough that the renormalized $\kappa$ is not yet driven to $0$ by anharmonic fluctuation effects. 
This implies that  the membrane can avoid crumpling if some new physics beyond the purely elastic model (\ref{free energy}) intervenes on some new length scale  $L_{\rm new}$  smaller than 
the orientational correlation length $\xi$, which is given by 
\beq 
\xi=\Lambda^{-1}e^{\ell_v} \ ,
\label{xi1}
\eeq
with $\ell_v$ defined as the RG ``time" at which $\kappa(\ell_v)=0$. (Here we have used the usual relation $L(\ell)=\Lambda^{-1}e^{\ell}$ between renormalization group ``time" $\ell$ and length scale $L$.)

 We can therefore calculate the maximum length  
$L_c$ that a membrane can have while still remaining uncrumpled by calculating the orientational correlation length $\xi$, and the ``new physics"
length scale $L_{\rm new}$ (which will depend on the membrane length $L_m$), and then equating the two.

We will begin by calculating $\xi$ in the limit  $\chi^2\rightarrow\chi_{_U}^2$ from below using the recursion relations (\ref{flowc}).  In this limit, since the bare parameters $g^0_1\propto 1/\kappa_0^2$ and $g^0_2\propto 1/\kappa_0^3$,  $g^0_2$  diverges faster than $g^0_1$ as $\kappa_0\rightarrow 0$;
 hence,  $g^0_2\gg g^0_1$ as  $\chi^2\rightarrow\chi_{_U}^2$ from below.  The recursion relation (\ref{ratio}) for 
the ratio ${g_2(\ell) \over g_1(\ell)}$ { then implies that in this region of parameter space, }$g_2(\ell)\gg g_1(\ell)$ for all $\ell$, since their 
ratio grows everywhere above the separatrix.  The value  $\ell_v$ of $\ell$ at which $\kappa(\ell)$ vanishes is the same as the value of $\ell$ at which $g_{2}(\ell)$ diverges, since, as can be seen from its definition,  $g_2(\ell)\rightarrow\infty$ as $\kappa(\ell)\rightarrow 0$. This value can be approximated, for small bare $g_{2}^0$,  by $\ell_1$, the value of $\ell$ at which $g_{2}(\ell=\ell_1)=\cO(1)$, since, once $g_2(\ell)$ gets to be of $\cO(1)$, only a finite, $\cO(1)$ further renormalization group time $\delta\ell$ is required for $g_2(\ell)$ to grow from $\cO(1)$ to $\infty$. This statement can be verified directly from the recursion relation (\ref{flow2}) for $g_2$ which, in the limit $g_1(\ell)\ll g_2(\ell)$ (a limit which we showed above  holds for all $\ell$ in the limit $\chi^2\rightarrow\chi_{_U}^2$ from below), can be solved analytically, yielding 
\beq\ell={1\over\epsilon}\ln\left[\left({g_2(\ell)\over \epsilon+15g_2(\ell)/2}\right)\left({\epsilon+15g_2^0/2\over g_2^0}\right)\right] \ .
\label{ell1inf}
\eeq
Evaluating this for the physical case $\epsilon=2$ in the limit $g_2^0\ll1$  with $g_2(\ell_1) =1$  gives
\beq
\ell_1={1\over2}\left[\ln\left({2\over g_2^0}\right)-\ln (19/2)\right]\,
\label{ell1}
\eeq
and 
\beq
\ell_v={1\over2}\ln\left({2\over g_2^0}\right) \,,
\label{ellinf}
\eeq
where in (\ref{ellinf}) we have used the fact that  $g_2(\ell_v)=\infty$, since $\kappa(\ell_v)=0$, by definition. As claimed, these two values of $\ell$ differ by an amount of $\cO(1)$. Of course, we do not actually know the precise value of this $\cO(1)$ constant, since our recursion relations
(\ref{flow1}) and (\ref{flow2}) are not accurate for $g_2\gg1$. However, that the difference $\ell_0-\ell_1$  {\it is} of $\cO(1)$ is clear, provided that the flows do not pass too close to the putative strong coupling fixed point in figure (\ref{fp}).

Using the result (\ref{ellinf}) for $\ell_v$ in our expression (\ref{xi1}) for the orientational correlation length gives
\beq
\xi=\Lambda^{-1}\sqrt{1\over g^0_2}\times O(1)={1\over B_0}\sqrt{\kappa_0^3\over \kbt}\times O(1) \ ,
\label{Linf1}
\eeq
where in the second equality we have used our expression (\ref{gdef}) for  $g_2$,  evaluated with the bare values $B_0$ and $\kappa_0$ of the parameters $B$ and $\kappa$, to evaluate $g_2^0$.

 We now turn to the calculation of the length scale $L_{\rm new}$ beyond which new physics  not included in  the elastic model (\ref{free energy}) can intervene  to prevent crumpling before this length scale is reached. We have already discussed one such piece of physics: self-avoidance. The associated  length scale $L_H$ is the typical distance  between successive contacts between neighboring turns of the spiral, and can cut off any tendency to crumpling in the spiral sections of the membrane. But as inspection of  Fig.~(\ref{dspiral}) makes clear, there is one section of the membrane for which this cutoff cannot work: the straight section connecting the two oppositely returning spirals.  This section has no neighbors, because it lies outside both spirals. It is therefore the section of the membrane that will crumple first, thereby inducing crumpling of the rest of the membrane. 

This straight, ``connecting" section of the membrane is stabilized by  surface tension, which arises because that section of the membrane could lower its energy by ``rolling up" into one or the other of the spiral sections it connects (since it should thereby be closer to the optimal spontaneous curvature).  It is not rolled up, of course, because the other spiral 
pulls it equally hard in the opposite direction. These two pulls create a non-zero surface tension $\sigma$, whose magnitude should be comparable to the Helfrich interaction in the outermost turn of the spiral, since it is the balance between that interaction, which works to open the spiral, and the spontaneous curvature term, which  tightens, that sets the scale of the energy of those outermost turns of the membrane, and, hence, the surface tension.

 Since we want $L_H<\xi$, we must determine $L_H$ using harmonic elastic theory, rather than the anharmonic elastic theory we used in our earlier discussion of the spiral state. This is so because $\xi$ is of order the length scale on which the renormalized $g_2(\ell)=\cO(1)$, since $\ell_0-\ell_1=\cO(1)$. Hence, on length scales $L_H<\xi$, $g_2\ll 1$, which implies that anharmonic effects are unimportant on these length scales.  (Recall that $g_1(\ell)\ll g_2(\ell)$ in this regime, so if 
$g_2(\ell)\ll 1$, $g_1(\ell)\ll 1$ as well.)

Since  at harmonic order we can ignore the Gaussian curvature terms in the free energy (\ref{free energy_h}), which are anharmonic, the free energy (\ref{free energy_h_RS}) becomes identical to that for a {\it fluid} membrane.  Therefore, the relation between $L_H$ and the spacing $d$ between successive turns of the membrane is the same as that in a lamellar phase of symmetric fluid membranes; that relation has long been known\cite{helfrich1978} to be 
\beq 
L_H=\sqrt{\ko\over\kbt}d \ .
\label{Lbump}
\eeq
To relate this $L_H$ to the total length $L_m$ of the membrane,  we first need to determine the shape of the spiral in the regime in which harmonic elastic theory is valid. This analysis is virtually identical to that done earlier for the anharmonic theory; the only modification is that the Helfrich potential is now~\cite{helfrich1978}

\bea
\mathcal{U}_{H_{\rm{h}}}(d)={\kbt\over L_H^2} ={(k_BT)^2\over \kappa_0d^2}\times\cO(1)
\ . \nonumber\\
\label{helharm}
\eea
As in our earlier treatment in section (\ref{L gt Lfree}) of the form of the spiral in the stable region of parameter space, balancing this Helfrich repulsion against the spontaneous curvature term   $C$ gives the form of the spiral. The reasoning is identical to that leading from equation (\ref{energyhel}) to equations {(\ref{ralpha2}) and (\ref{alpha}) of section (\ref{L gt Lfree}), but with $\eta$ everywhere replaced by $0$. This leads easily to
\bea 
r=R_{0h}^{1/3} s^{2/3} \ ,
\label{ralphah} 
\eea
where 
\bea
R_{0h}=\left({k_B T\over\kappa_0}\right)^2\frac{ \kappa_0}{C}\times\cO(1)
\ .
\label{R_0h}
\eea
Combining this result (\ref{ralphah}) for the shape of the spiral with the relation $d(s)\sim r{\dd r\over\dd s}$ gives for the spacing between successive turns of the membrane:
\beq
d(s)\sim R_{0h}^{2/3}s^{1/3} \ ,
\label{ds}
\eeq
The largest value of this is at the outer edge of the membrane, where $s=L_m/2$, which implies 
\beq
d_{\rm{max}}\sim R_{0h}^{2/3}L_m^{1/3} \ .
\label{dmax}
\eeq
Using this value of $d_{\rm max}$ in our expression (\ref{helharm}) for the Helfrich interaction $\mathcal{U}_{H_h}(d_{\rm max})$, and estimating the surface tension $\sigma\sim\mathcal{U}_{H_h}(d_{\rm max})$ gives
\beq
\sigma\sim\mathcal{U}_{H}(d_{\rm max}) ={(k_BT)^2\over 
\kappa_0d_{\rm{max}}^2}\sim {C^{4/3}\kappa_0^{1/3}\over (k_BT)^{2/3}L_m^{2/3}} \ ,
\label{sigma}
\eeq
where $L_m$ is the linear extent of the membrane. 

Associated with this surface tension is the "new physics" length scale  $L_{\rm new}$  we seek: namely, the length scale $L_\sigma$ at which the surface tension energy becomes comparable to the bending energy. At this scale, the surface tension energy $\sigma A$ should be comparable to the bending energy ${\kappa_0\over L_\sigma^2} A$, where $A$ is the area. Equating these and solving for $L_\sigma$ gives
\beq
L_\sigma=\sqrt{\kappa\over\sigma} \sim\left({k_BT
\kappa_0L_m\over C^2}\right)^{1/3}\ .
\label{Lsig}
\eeq
Equating this to $\xi$ and solving for $L_m$ gives the maximum size $L_c$ of the membrane that can be stable:
\beq 
L_c\sim{\kappa_0^{7/2}C_0^2\over (k_BT)^{5/2}B_0^3}\propto (\chi_{_U}^2 - \chi^2)^{7/2}  \ ,\label{L_c}
\eeq
where the dependence on $\chi$ follows from our expression (\ref{kappa}) for the dependence of the bare bending stiffness $\kappa_0$ on $\chi$.

This expression, and the scaling law $L_c\propto (\chi_{_U}^2 - \chi^2)^{7/2}$, will break down if $\chi^2$ gets too close to $\chi_{_U}^2$, since then $L_c$ gets too small for our long-wavelength approach to be valid. However, because of the $T^{-{5\over2}}$ dependence of $L_c$ on temperature, the range of $\chi_{_U}^2 - \chi^2$ 
 over which the scaling law will be valid will get quite large if the temperature $T$ is small (in particular, for $\kbt\ll\kappa$).

Note that the result (\ref{L_c}) will also break down as $\chi^2\to\chi_{_{_L}}^2$ from above, because then the flows will pass close to the putative strong coupling fixed point in figure (\ref{fp}). Since we know nothing quantitative about that fixed point, we can say nothing quantitative about $L_c$ in this limit, except that it must diverge, since the flows will linger for a large renormalization group time near that fixed point (since it {\it is} a fixed point), which means that the orientational correlation length $\xi$ must diverge as $\chi^2\to\chi_{_{_L}}^2$ from above.  Readers who prefer a perturbation theory argument for divergence of $\xi$ as  $\chi^2\to\chi_{_{_L}}^2$ from above to this RG approach can find one in Appendix (\ref{vanishkapp}).

The above argument leading to equation (\ref{L_c}) for $L_c$ assumed, as discussed earlier, that the straight part of the membrane connecting the two spirals will crumple first. To verify this, we must show  that $L_\sigma\gg L_H^{\rm{max}}$, because then, as we increase membrane size, the straight section will crumple (because 
$L_\sigma$ has exceeded $\xi$) when the curled up section is still uncrumpled  (because 
$L_H^{\rm{max}}$ has not yet exceeded $\xi$).

To demonstrate this, we simply need to take the ratio of $L_\sigma$, as given by equation (\ref{Lsig}), to $L_H^{\rm{max}}$, as given by equation (\ref{Lbump}), with $d=d_{\rm max}$. 
This gives
\beq 
L_H^{\rm{max}}\sim \sqrt{\frac{\kappa_0}{k_BT}}R_{0h}^{2/3}L_m^{1/3} \ .
\label{lbmax}
\eeq
Taking the ratio of  $L_\sigma$ to this $L_H^{\rm{max}}$ gives
\beq
{L_\sigma\over L_H^{\rm{max}}}\sim\sqrt{\kappa_0\over k_BT}\gg1 \ ,
\label{Ls/Lb}
\eeq
where the last strong inequality will hold at low temperatures $\kbt\ll\kappa_0$, which is the condition for the validity of all of the above arguments in any case.

  See Figs.~\ref{k-chi} for schematic phase diagrams in 
the $\chi- \kappa^\prime$ and $\chi^2-L$ planes.

\section{Summary}\label{sum}

We have here developed an elastic theory  for asymmetric tethered membranes,  and used it to study their
statistical mechanics. 
Our theory includes  a coupling between local in-plane lattice dilations and membrane curvature  that is  forbidden by symmetry in inversions-symmetric tethered membranes. 
When   this coupling  is sufficiently weak,   it causes asymmetric membranes to have a completely different structure from symmetric membranes: rather than being flat, asymmetric membranes curl up into a ``double spiral" structure, as illustrated in Fig.~\ref{dspiral}. The shape of this spiral is universal, and characterized by scaling exponents which can all be related to the anomalous elastic exponent $\eta$ for bending elasticity in {\it symmetric} membranes. 
For stronger dilation-dependence, the membrane crumples. Thus structural (inversion) asymmetry provides a new route to crumpling of tethered membranes.  This inversion-asymmetry induced crumpling can happen in two ways:

\noindent 1) for the strongest dilation-dependence, the membrane crumples no matter how small it is. 

\noindent 2) for intermediate dilation-dependence, membranes only crumple if their size exceeds a critical threshold. 

These results are summarized in the phase diagrams (\ref{k-chi}).

At temperature $T=0$, 
the spiral state of an asymmetric tethered membrane remains smooth and 
necessarily bends in only one direction. The shape of a cross-section in the plane of this bent direction is a double 
spiral composed of two Archimedes' spirals and a straight section joining them. 
The reason the membrane bends only in one direction is that bending along both the directions 
would generate Gaussian curvature, resulting into free energy costs that diverge in 
the thermodynamic limit. 

For $T>0$,  this unidirectionally bent double-spiral structure persists,
but the double spiral is now considerably swelled up, with a structure now given by equation (\ref{rtheta}), with a universal exponent $\nu$ which we can relate to the anomalous bend elasticity exponent $\eta$ of {\it symmetric} membranes. This swelling arises
from the competition between the Helfrich interactions between the successive layers 
in each of the spirals and the spontaneous curvature. 
This phase shows long range orientational order, and is the analog of the statistically flat phase of 
symmetric tethered membranes at finite $T$.  For a rectangular membrane, the free energy is lowest if the  membrane rolls up along the longer axis   (as opposed to  rolling up along the shorter axis). Interestingly, however, the spiral state formed by rolling up along the shorter axis is a {\em long-lived}, metastable state, with a life time that diverges in the thermodynamic limit.

}

In addition to the long range interactions between the local Gaussian curvatures 
present in symmetric tethered membranes, asymmetric membranes exhibit long range interactions between the 
local Gaussian and mean curvatures.

  This theory can be   tested in numerical simulations,  and controlled experiments on  a variety of membrane systems, including: graphene coated by some substance on one side, artificial deposits of spectrin filaments on model lipid membranes, a bilayer made of a usual lipid monolayer and a symmetric tethered membrane, and in-vitro experiments on red blood cell membrane extracts.

\section{Acknowledgements}
T.B. and A.B.  thank the Alexander von
Humboldt Stiftung (Germany) for partial financial support under the Research 
Group Linkage Programme scheme (2016). 
T.B. and J. T. thank the Max-Planck Institut f\"ur Physik Komplexer Systeme, 
Dresden, Germany, for their hospitality and financial support while this
 work was underway.

 \section*{Appendix I: Glossary}
 \label{gloss}

 In this glossary, we list, and give rough definitions for, all of the symbols used in this paper, in the order in which they appear. We also  give the equation that precisely defines them, which is not, in all cases, the first equation in which the symbol appears.

 \begin{enumerate}
 	\item $a$: Thickness of the membrane (Eq. \ref{rspiral}).
  \item $r_0$: Radius of the central hole of the spiral membrane (Eq. \ref{rspiral}).
    \item $\kappa_0$: Bare bend modulus of an asymmetric tethered membrane {\em after} integrating out the in-plane elastic modes (Eq. \ref{kappa}).
  \item $C$: Phenomenological coefficient that determines the free energy gain due to spontaneous curvature of an asymmetric tethered membrane (Eq. \ref{free energy}).
    \item $\kappa(\bq)$: Renormalized scale-dependent bend modulus of an asymmetric tethered membrane (Eq. \ref{kappaq}).
  \item $\eta$: Universal scaling exponent  for the divergence of $\kappa(\bq)$ in the infra-red ($\bq\to{\bf 0}$ limit in symmetric tethered membranes (Eq. \ref{kappaq}).
  \item $\mu_0,\,\lambda_0$: Bare Lam\'{e} coefficients  for the in-plane elasticity of the membrane.  (Eq. \ref{free energy}).
  \item $\chi$: Symmetry-permitted coefficient coupling  dilation and mean curvature (Eq. \ref{free energy}). 
  \item $\chi_{_U}^2$: Upper limit on $\chi^2$, such that for $\chi^2>\chi_{_U}^2$, an asymmetric tethered membrane of any size, however small, necessarily crumples even at $T=0$ (Eq. \ref{L_c11}).
  \item $L_c$: Critical linear size of the membrane above which the membrane crumples (Eq. \ref{L_c11}).
 \item $\kappa'$: Bare bend modulus of a tethered membrane  in the absence of lattice dilation.  (Eq. \ref{free energy}).
 \item $P_{ij}$: transverse projection operator (Eq. \ref{Pdef}).
 \item $A$: Coefficient of the inversion-symmetric nonlinear term in the effective free energy after integrating out the in-plane elastic modes (Eq. \ref{free energy_h_FT}).
  \item $B$: Coefficient of the inversion-asymmetric nonlinear term in the effective free energy after integrating out the in-plane elastic modes (Eq. \ref{free energy_h_FT}).
  \item $g_1$: Dimensionless effective coupling constant, also present in symmetric tethered membranes.
  (Eq. \ref{gdef}).
  \item $g_2$: Dimensionless effective coupling constant; not present in symmetric tethered membranes.
  (Eq. \ref{gdef}).
  \item $\epsilon = 4-D$ is the small parameter in perturbative RG employed here (Eq. \ref{gdef}).
  \item $\ell$: Renormalization group ''time`` with $\exp(\ell)a$ being an associated length. 
  (Eqs. (\ref{renor_kappa},\ref{renor_A},\ref{renor_B})).
  \item $\xi_{NL}$: The non-linear length $\xi_{_{NL}}$ is the length scale at which $\kappa$ starts to acquire appreciable fluctuation corrections (Eq. \ref{kappaq}).
  \item $\eta$: Universal scaling exponent that controls the divergence of $\kappa(q)$ in the infra-red limit in asymmetric tethered membranes in their stable spiral phase. (Eq. \ref{kappaq}).
  \item $L_H$: Typical distance between  points of contact between two successive turns \sout{in} of a spiral membrane.
  (Eq. \ref{lcn}).
  \item ${\mathcal U}_H$: Helfrich interaction between the successive layers in a stack of membranes 
  (Eq. \ref{hel}).
  \item $d$: Typical distance between the two  successive turns of the spiral membrane (Eq. \ref{ls}).
  \item $D$: Internal space dimension of the membrane; $D=2$ is the physical case. Thus the physical embedding dimension is $D+1=3$ (Eq. \ref{Ddef}).
  \item $\chi_{_{_L}}^2$: Lower limit on $\chi^2$, such that for $\chi^2 < \chi_{_{_L}}^2$, only a membrane that is big enough ($L_m>L_c$) crumples (Eq. \ref{chiL3}).
  \item $\xi$: Orientational correlation length (Eq. \ref{xi1}). 
  \item $L_m$: Linear size of the membrane (Eq. \ref{dmax}).
  \item $\sigma$: Surface tension of the segment of the membrane that joins the two spirals in the spiral phase 
  (Eq. \ref{sigma}).
  \item $r_{max}$: Radius of the outermost turn of the spiral in the spiral phase (Eq. \ref{rmax}).
     \end{enumerate}

 \section*{Appendix II: Parameter estimates}
\label{par}

The following table gives the values of $\kappa_0$ for a variety of lipids. These are taken from~\cite{kappa-vals}.

\begin{table}[h!]
\begin{center}
\begin{tabular}{ |p{3cm}|p{3cm}| }
 \hline
 \multicolumn{2}{|c|}{Lipid bending rigidity} \\
 \hline
Lipid & ${\kappa_0}$ (erg)  \\
 \hline
DMPC & $(1.15\pm 0.15) \times 10^{-12}$ \\
DMPC + 20\% Cholesterol & $(2.1\pm 0.25) \times 10^{-12}$ \\
DMPC + 30\% Cholesterol & $(4.0\pm 0.8) \times 10^{-12}$ \\
egg-PC & $(1.15\pm 0.15) \times 10^{-12}$ \\
DMPC + C5-PC 1:1 & $(1.7\pm 0.2)\times 10^{-12}$ \\
G-DG & $(1.5-4)\times 10^{-13}$ \\
Erythrocyte & $(3-7)\times 10^{-13}$ \\
\hline
\end{tabular}
\caption{{Table of} bending rigidities of some lipids.}
 \label{tab1}
\end{center}
\end{table}

\vskip0.5cm

The following table gives the values of the inverse spontaneous curvature ${C_0\over\kappa_0}$ for a variety of lipids.  These values are taken from~\cite{kamal2009}.

\begin{table}[h!]
\begin{center}
\begin{tabular}{ |p{3cm}|p{3cm}| }
 \hline
 \multicolumn{2}{|c|}{Lipid spontaneous curvature} \\
 \hline
Lipid & ${C/\kappa} (nm^-1)$ \\
\hline
L-lyso PC & $1/5.8$ \\
O-lyso PC & $1/3.8$ \\
P-lyso PC & $1/6.8$ \\
L-lyso PE & $<1/40$ \\
O-lyso PE & $<1/40$ \\
S-lyso PE & $<1/40$ \\
DOPS & $1/14.4$ \\
DOPC & $-1/20$ \\
PA & $-1/4.6$ \\
DOPE & $-1/3$ \\
Cholesterol	& $-1/2.9$ \\
DCG	& $-1/1.3$ \\
\hline
\end{tabular}
\caption{{Table of} inverse spontaneous curvature of lipids.}
 \label{tab2}
\end{center}
\end{table}

 \section*{Appendix III:  Rotationally invariant free energy }
 \label{full-rot}

We now formulate the full three-dimensional (3D) rotationally invariant free energy functional ${\cal F}_{rot}$ for asymmetric tethered membranes. To begin with we define a set of two internal orthogonal coordinates ${\boldsymbol \tilde \sigma} = (\tilde \sigma^1,\tilde\sigma^2)$ that can be used to define an intrinsic metric $g_{ab}$ via
 \begin{equation}
  g_{ab}=\partial_a {\bf R}({\boldsymbol\tilde\sigma})\cdot \partial_b {\bf R} ({\boldsymbol\tilde\sigma}),
 \end{equation}
where $\bf R$ is a 3d position vector; $a,b=1,2$ are the labels of $\boldsymbol\tilde \sigma$. We define the coordinates in a way such that the minimum free energy configuration corresponds to $g_{ab}=\delta_{ab}$, the unit matrix. We further define the strain tensor $u_{ab}$ for a deformed configuration as
\begin{equation}
 u_{ab}=g_{ab}-\delta_{ab}.
\end{equation}
 The full rotationally invariant free energy functional ${\mathcal F}_{rot}$ takes the generic form
\begin{equation}
 {\mathcal F}_{rot}=\frac{1}{2}\int d^2 \tilde\sigma [ \frac{\kappa^\prime}{2} ({\rm Tr}\, {\bf K})^2 + \mu u^{ab} u_{ab} + \frac{\lambda}{2} u^a_a\,u^b_b+V(u_{aa}) {\rm Tr}\, {\bf K}],\label{frot}
\end{equation}
where $V (u_{aa})$ is a general function of the bulk strain $u_{aa}$. In (\ref{frot}), we have kept only terms that prove to be relevant at long wavelengths. Here, $\bf K$ is the curvature tensor, and $\rm Tr$   denotes the trace of a matrix. Since $\bf K$ is odd under inversion of $\bf R$, the last term in (\ref{frot}) breaks symmetry under parity inversion, and  therefore cannot be present in inversion-symmetric membranes. 

 For small deformations, we Taylor-expand $V(u_{aa})$ in powers of $u_{aa}$ to write
\begin{equation}
 V(u_{aa})=C+\chi u_{aa} \,,\label{Vform}
\end{equation}
to linear order in $u_{aa}$, which proves to be the highest order to which we need go to include all relevant terms.

For a nearly flat segment, it is convenient to use the Monge gauge,   in which ${\bf R} = (x,y,h(x,y))$, with  the orthogonal coordinates $x,y$   as the internal coordinates. In the Monge gauge, (\ref{frot}) readily reduces to (\ref{free energy}) with $V$ as given in (\ref{Vform}). 
 
\section*{Appendix IV: RG calculation}
\label{RG}

 In this appendix we systematically evaluate the Feynman diagrams for $\kappa,\,A$ and $B$. First consider the diagrams for $\kappa$ as shown in Fig.~\ref{kappa_dia-all}. Diagram  \ref{kappa_dia-all}(a) { gives a contribution $(\delta H)_{5a}$ to the renormalized Hamiltonian 
\begin{widetext}
\begin{equation}
(\delta H)_{5a}=\sum_\bk|h(\bk)|^2\left(\frac{\kbt A k^4}{2\kappa}\int_> \frac{d^D q}{(2\pi)^D} \frac{1}{|{\bf k -q}|^4} [k_jP_{ij}({\bf q})k_n P_{mn}({\bf q})]\right)=\sum_\bk k^4|h(\bk)|^2\left(\frac{\kbt A}{2\kappa}\frac{(D-1)(D+1)}{D(D+2)}\int_> \frac{d^Dq}{(2\pi)^D}\frac{1} 
 {q^4}\right) \,,
\end{equation}
\end{widetext}
where in the second equality we have worked to leading order in the external wavevector $\bk$. Here $\int_>$ denotes an integral over the momentum shell $b^{-1}\Lambda<|\bq|<\Lambda$, where $\Lambda$ is the ultraviolet cutoff. The proportionality of this correction to  $\sum_\bk k^4|h(\bk)|^2$ identifies it as a correction to $\kappa$, and implies
\begin{equation}
(\delta \kappa)_{5a}=\frac{\kbt A}{\kappa}\frac{(D-1)(D+1)}{D(D+2)}\int_> \frac{d^Dq}{(2\pi)^D}\frac{1} 
 {q^4} \,.
\end{equation}
}

Diagram \ref{kappa_dia-all}(b)   {  likewise gives a contribution $(\delta H)_{5b}$ to the renormalized Hamiltonian 
\begin{widetext}
\begin{equation}
(\delta H)_{5b}=-\sum_\bk|h(\bk)|^2 \left(\frac{B^2}{2\kappa^2}\int_>\frac{d^Dq}{(2\pi)^D}k_i k_j k_m k_n \frac{P_{ij}({\bf q})P_{mn}({\bf q}) }{q^4}\right) = \sum_\bk k^4|h(\bk)|^2\left(-\frac{B^2}{2\kappa^2}\frac{(D-1)(D+1)}{D(D+2)}\int_> \frac{d^Dq}{(2\pi)^D} \frac{1}{q^4}\right) \,,
\end{equation}
\end{widetext}
which can, as we just did for \ref{kappa_dia-all}(a), be interpreted as a renormalization of $\kappa$:
\begin{equation}
(\delta \kappa)_{5b}=-\frac{B^2}{\kappa^2}\frac{(D-1)(D+1)}{D(D+2)}\int_> \frac{d^Dq}{(2\pi)^D} \frac{1}{q^4} \,.
\end{equation}}

Similarly, diagram \ref{kappa_dia-all}(c)  { contributes 
\begin{widetext}
\begin{equation}
(\delta H)_{5c}=-\sum_\bk|h(\bk)|^2 \left(\frac{B^2}{2\kappa^2}k_ik_mk_n\int_>\frac{d^Dq}{(2\pi)^D}\frac{q_jP_{ij}({\bf k-q}) P_{mn}({\bf q})} { q^4}\right) = \sum_\bk|h(\bk)|^2 \left(-\frac{B^2k^4}{2\kappa^2}\frac{(D-1)(D+1)}{D(D+2)}\int_> \frac{d^Dq}{(2\pi)^D} \frac{1}{q^4}\right) \,,
\end{equation}
\end{widetext}
which is the} same as diagram \ref{kappa_dia-all}(b).  { Hence
\begin{equation}
(\delta \kappa)_{5c}=(\delta \kappa)_{5b}=-\frac{B^2}{\kappa^2}\frac{(D-1)(D+1)}{D(D+2)}\int_> \frac{d^Dq}{(2\pi)^D} \frac{1}{q^4} \,.
\end{equation}}

The last diagram \ref{kappa_dia-all}(d) { contributes
\begin{widetext}
\begin{equation}
(\delta H)_{5d}=-\sum_\bk|h(\bk)|^2 \left( \frac{B^2k^4}{4\kappa^2} P_{ij}({\bf k})P_{mn}({\bf k}) \int_> \frac{d^Dq}{(2\pi)^D} \frac{q_iq_jq_mq_n}{q^8}\right)=\sum_\bk|h(\bk)|^2 \left(-\frac{B^2k^4}{4\kappa^2}\frac{(D-1)(D+2)}{D(D+2)}\int_>\frac{d^Dq}{(2\pi)^D}\frac{1}{q^4}\right) \,.
\end{equation}
\end{widetext}
This implies a correction to $\kappa$ given by
\begin{equation}
(\delta \kappa)_{5d}=-\frac{B^2}{2\kappa^2}\frac{(D-1)(D+1)}{D(D+2)}\int_> \frac{d^Dq}{(2\pi)^D} \frac{1}{q^4} \,.
\end{equation}}

%

This completes the graphs for $\kappa$. 
We now  turn to the graphs for $A$. The only non-zero graph that contributes to $A$ at one loop order is Fig.~\ref{Adiag}, { which contributes to the renormalized Hamiltonian a term}: 
\begin{widetext}
\begin{eqnarray}
(\delta H)_{6}&=& \sum_\bk \left|P_{ij}(\bk) A_{ij}(\bk)\right|^2\left(- \frac{A^2\kbt}{16\kappa^2} \int_> \frac{d^Dq}{(2\pi)^D}  \frac{q_mq_lq_nq_p} {q^8}\right)\nonumber\\
 &=& \sum_\bk \left|P_{ij}(\bk) A_{ij}(\bk)\right|^2\left(-\frac{A^2}{16\kappa^2}\frac{(D-1)(D+1)}{D(D+2)}\int_> \frac{d^Dq}{(2\pi)^D}\frac{1}{q^4}\right) \,.
\end{eqnarray}
\end{widetext}
The structure of this term identifies it as a renormalization of $A$ given by
\begin{equation}
(\delta A)_{6}=-\frac{A^2}{2\kappa^2}\frac{(D-1)(D+1)}{D(D+2)}\int_> \frac{d^Dq}{(2\pi)^D} \frac{1}{q^4} \,.
\end{equation}
Similarly, the one-loop diagram for $B$ in Fig.~\ref{Bdiag} is
\begin{widetext}
\begin{equation}
 \frac{AB}{4\kappa^2} P_{mn}({\bf k})P_{lp}({\bf k})\int_>\frac{d^Dq}{(2\pi)^D} \frac{q_mq_nq_lq_p} {q^8}=\frac{AB}{4\kappa^2}\frac{(D-1)(D+1)}{D(D+2)})\int_> \frac{d^Dq}{(2\pi)^D} \frac{1}{q^4}.
\end{equation}
\end{widetext}

Combining all of these corrections and performing the spatial  and field rescalings described in (\ref{RGMT}) leads to the following discrete recursion relations for $\kappa,\,A$ and $B$:

\begin{eqnarray} \kappa^{\prime}&=&b^{-\eta}\kappa[1+ (\frac{A K_D}{\kappa^2}- \frac{5 B^2 K_D}{2\kappa^3}) \int_{b^{-1}\Lambda}^\Lambda d q\,q^{D-5}],\nonumber \\  A^{\prime}&=& b^{-2\eta+4-D}A[1-\frac{A K_D}{2\kappa^2} { \int_{b^{-1}\Lambda}^\Lambda d q\,q^{D-5}}],\nonumber \\ B^{\prime}&=& b^{\frac{-3\eta+4-D}{2}} B[1-\frac{A K_D}{2 \kappa^2} { \int_{b^{-1}\Lambda}^\Lambda d q\,q^{D-5}}] \,, \end{eqnarray}
 where we have defined $K_D=\frac{(D^2-1)S_D}{(2\pi)^DD(D+2)}$, with $S_D$ is the surface hyper-area of a D-dimensional sphere of unit radius.

Taking $b=1+d\ell$ with $\ell$ differential enables us to rewrite these recursion relations in the usual way as differential equations; the result is the differential recursion relations given in (\ref{RGMT}).  

 We now turn to Feynmann diagrams that look as though they might contribute, but which in fact do not.
 

\section*{Appendix V: Feynman diagrams that vanish in the long wavelength limit}\label{vanishing diagrams}

We now evaluate some of the Feynman diagrams that  look as though they might renormalize various parameters of our model, but which actually make no contributions to them. Rather, these graphs only renormalize higher order coefficients that are irrelevant in the long wavelength limit. 

\begin{figure}
\includegraphics[width=8cm]{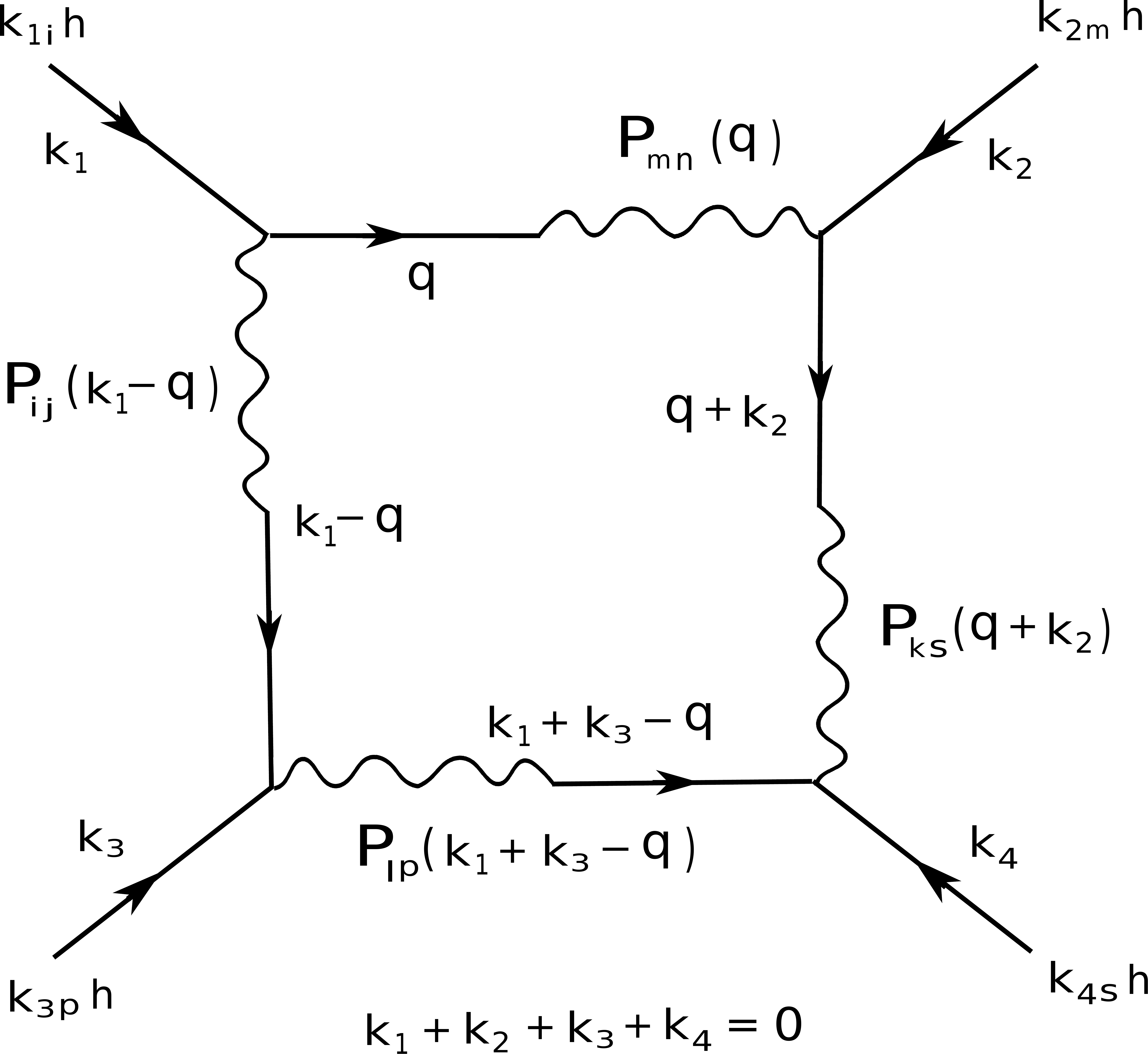}
 \caption{ A one-loop Feynman diagram that  appears to correct $A$,  but which in fact vanishes in the limit of zero external wavevectors.} \label{vanisha}
 \end{figure}

Consider the one-loop diagram in Fig.~\ref{vanisha},  which looks as though it could contribute to  $A$,  since it is proportional to $h^4$. We will now show that this graph  actually makes a negligible contribution to the renormalized Hamiltonian in the limit of  small external wavevectors.

This graph contributes to the Hamiltonian a term 
\begin{widetext}
\begin{eqnarray}
(\delta H)_{11}&&\propto{ \sum_{\bk_1,\bk_2,\bk_3,\bk_4}\delta^K_{\bk_1+\bk_2+\bk_3+\bk_4}}k_{1i}k_{2m}k_{3p}k_{4s} h(\bk_1)h(\bk_2)h(\bk_3)h(\bk_4) \nonumber\\&&\times(\int d^Dq P_{ij}({\bf k}_1-{\bf q})P_{mn}({\bf q})P_{ks}({\bf k}_2+{\bf q}) P_{lp}({\bf k}_1 + {\bf k}_3-{\bf q})
|{\bf k}_1 -{\bf q}|^2 
|{\bf k}_1+{\bf k}_3-{\bf q}|^2 |{\bf q}+{\bf k}_2|^2 \nonumber\\&&({\bf k}_1 - {\bf q})_l ({\bf k}_2+{\bf q})_nq_j({\bf k}_1 + {\bf k}_3 - {\bf q})_k
\langle |h({\bf q})|^2\rangle \langle |h({\bf k}_2+{\bf q})|^2\rangle\langle |h({\bf k}_1-{\bf q})|^2\rangle \langle |h({\bf k}_1 + {\bf k}_3 -{\bf q})|^2\rangle),
\end{eqnarray}
\end{widetext}

which is irrelevant in the limit ${\bf k}_1,\,{\bf k}_2,\,{\bf k}_3\rightarrow 0$.  This may be seen as follows: projection operator $P_{ks}({\bf k}_2+\bq)$ kills $({\bf k}_1 + {\bf k}_3 - {\bf q})_k$ as 
${\bf k}_1,\,{\bf k}_2,\,{\bf k}_3\rightarrow 0$, because in that limit it becomes $-P_{ks}(\bq)q_k$, which is the $s$ component of the projection of $\bq$ orthogonal to itself.  This projection obviously vanishes.
Therefore, this graph can only generate terms  schematically of $\cO(k^6)h^4$, where $\bf k$ is any of the external wavevectors.  Such terms have two more powers of $k$ than the $A$ term we already have in the Hamiltonioan, and are, hence,  are irrelevant as $\bk\to0$.

 Similarly, there are one-loop diagrams that appear to correct $B$,  but actually yield irrelevant contributions in the limit of zero external wavevector. A representative of such diagrams is shown in Fig.~\ref{vanishb}. It  contributes to the Hamiltonian a term
\begin{widetext}
\begin{eqnarray}
(\delta H)_{12}&&\propto{ \sum_{\bq_1,\bq_2,\bq_3}\delta^K_{\bq_1+\bq_2+\bq_3}}q_{1m}q_{2n}|\bq_3|^2P_{ij}(\bq_3) h(\bq_1)h(\bq_2)h(\bq_3) \nonumber\\&&\times\int d^Dq P_{ms}({\bf q}+{\bf q}_1 +{\bf q}_3) P_{ns} ({\bf q}+{\bf q}_2) q_sq_p ({\bf q} + {\bf q}_2)_j ({\bf q}+{\bf q}_1+{\bf q}_3)_i
\langle |h({\bf q}+{\bf q}_1 +{\bf q}_3)|^2\rangle \langle |h({\bf q}+{\bf q}_2)|^2\rangle \langle |h({\bf q}|^2\rangle \,.\nonumber\\
\end{eqnarray}
Once again, as the external momenta go to zero, the $P_{ns}$ projection operator kills the $q_s$, and this graph vanishes.
\begin{figure}
 \includegraphics[width=10cm]{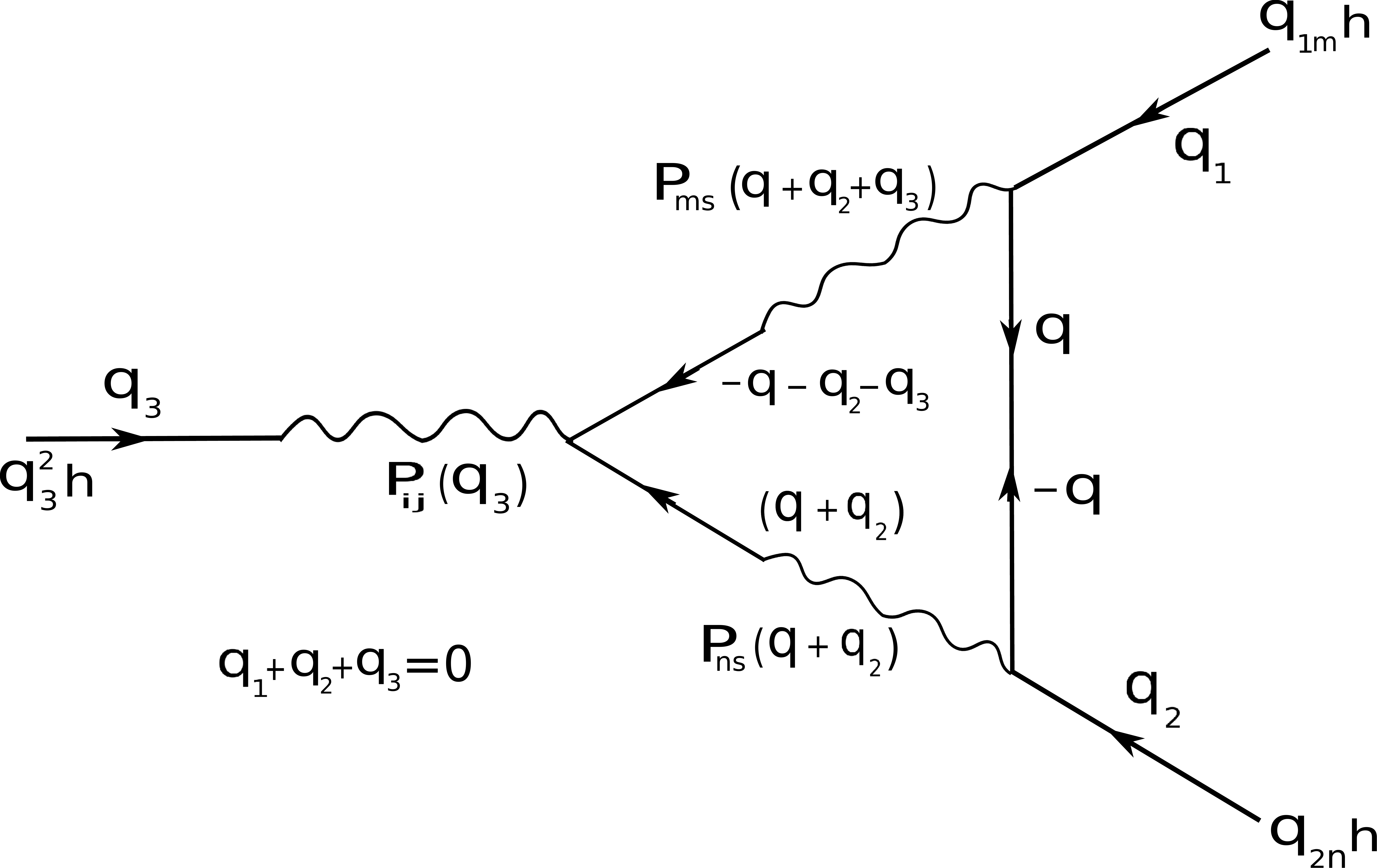}
 \caption{ A typical one-loop diagram that corrects $B$ that vanishes in the limit of zero external wavevectors.} \label{vanishb}
\end{figure}
\end{widetext}

\section*{Appendix VI: Orientational correlation length $\xi$ in  lowest order perturbation theory for $\kappa$}\label{vanishkapp}

We now show that the  upper and lower limits $\chi_U$ and $\chi_L$ on the asymmetry parameter $\chi$  can be extracted from the one-loop bare perturbation
theory for $\kappa$, based on the diagrams (\ref{kappa_dia-all}).  This calculation is the same as the RG calculation of Appendix blah, except that now we extend the integrals over wavevector down to an infrared cutoff $q_{\rm min}\equiv{2\pi\over L}$, where $L$ is the length scale on which we wish to know the effective value $\kappa_e$ of the bend modulus $\kappa$.   Evaluating the leading order (i.e., one loop) perturbative corrections to $\kappa$ coming from wavevectors $q<{2\pi/L}$.  We thereby obtain in $D=2$ 
\begin{equation}
 \kappa_e-\kappa_0 \approx \frac{3}{8}\left(\frac{A}{\kappa_0}
 -\frac{5 B^2}{4\kappa_0^2}\right) \int_{2\pi/L}^{\Lambda} \frac{d^2 q}{(2\pi)^2 
q^4}.\label{crumplekappa}
 \end{equation}
 
 { We are only interested here in the regime of parameter space in which $\kappa_0>0$ (indeed, that is the only regime in which this perturbative calculation even makes sense). In this regime,  if the factor in parentheses above is $>0$, $\kappa_e>\kappa_0>0$, so the orientational correlation length $\xi$ (which, we recall, is the length scale on which $\kappa_e$ vanishes) is infinite. Hence, for $\xi$ to be finite, we must have 
 \beq
 \frac{A}{\kappa_{0}}
 -\frac{5 B^2}{4\kappa_{ 0}^2}>0 \,,
 \label{chiL1}
 \eeq
 or, equivalently,
 \beq
 B^2>{4A\kappa_0\over5} \,.
 \label{chiL2}
 \eeq
 Using our expressions (\ref{kappa}) for $\kappa_0$, (\ref{Belasdef}) for $B$, and (\ref{Aelasdef}) for $A$ in this inequality, we can rewrite it as an inequality for the asymmetry parameter $\chi$: 
 }



\begin{equation}\label{chi_c eq}
\chi^2> \chi_{_{_L}}^2=\frac{A\kappa_0  
(2\mu+\lambda)^2}{A(2\mu+\lambda)+5\mu^2}=\frac{\kappa_0 
(2\mu+\lambda)}{1+\frac{5\mu}{4(\mu+\lambda)}} =\frac{\chi_U^2}{1+\frac{5\mu}{4(\mu+\lambda)}}\,.
\end{equation}
 Note that this has exactly the same form as our expression equation (\ref{chiL3})
for $\chi_L^2$ obtained from the RG, with the dimensionless $\cO(1)$ constant $\rho$ given by $\rho=5/4$. 
\\


\section*{Appendix VII: Vanishing of $\kappa$ on the unstable side of the separatrix}

We use the flow equation (\ref{renor_kappa}) for $\kappa$ and the definition (\ref{gdef}) of $g_2$,  and note that deep inside the crumpled phase, $g_2\gg g_1$.  This allows us to write
\begin{equation}
 \frac{d\kappa}{dl}\approx \kappa (-\eta - \frac{\mathfrak{G}B^2}{\kappa^3})\,,
 \label{kapparenorm2}
\end{equation}
where $\mathfrak{G}\equiv B^2K_Dk_BT\Lambda^{-\epsilon}$ is a constant under renormalization.
Further, the choice $\epsilon = 3\eta + 2 g_1$  makes the rhs of (\ref{renor_B}) vanish. With this choice and noting that $g_2>> \eta$, (\ref{kapparenorm2}) becomes
\begin{equation}
 \frac{d\kappa}{dl}=-\frac{\mathfrak{G}B^2}{\kappa^2} \,,
\end{equation}
with $B$ now also constant under renormalization. 
This yields
\begin{equation}
 \kappa^3(\ell) =  \kappa^3 (\ell=0) - \mathfrak{G}B^2 \ell,
 \end{equation}
indicating that $\kappa$ vanishes and become negative in a {\em finite} renormalization group time. As one approaches the separatrix from the crumpled phase, the  RG ``time" { required} for $\kappa$ to vanish is  becomes {ever} larger,  { ultimately diverging} on the separatrix.

\section*{Appendix VIII: Demonstration that the hole in the middle for $T=0$ is negligibly small}\label{hole1}

In this appendix, we show that the radius
$r_0$ of the hole in the middle of the spiral at $T=0$ is a negligible fraction of the total radius $R_T$ of the membrane. 
We consider a rectangular  membrane of dimensions $L_\parallel \times L_{ \perp}$, with $L_\parallel > L_{ \perp}$.  As discussed in the
main text, for this geometry the membrane will bend along the $L_{ \perp}$ direction. In our expression (\ref{rspiral}) for the spiral of Archimedes, we have 
chosen $\theta=0$ 
to be the innermost edge of the membrane. Hence $r_0$ is the size of the 
{\em hole} 
left in the center of the spiral. We can determine $r_0$ by minimizing the 
energy, which is given by
\bea
E=L_{||}\int_0^{L_\bot} ds \left[-{C \over R(s)} + {\kappa_0 \over {2R^2(s)}} 
\right], \label{energysp}
\eea 
where $R(s)$ is the radius of curvature of the membrane at distance $s$ along from the inner edge of the spiral.
This implies
\bea
{ds \over d\theta} = \sqrt{r^2+\left({dr\over d\theta}\right)^2}\approx r \,, \label{arclength}
\eea
 where the approximate equality holds if ${dr\over d\theta}\ll r$. For the spiral of Archimedes  (\ref{rspiral}), this will clearly always be true provided  that $r_0\gg a$,  since  then $r(\theta)\gg a$ for all 
$\theta$.  If $r_0$ is {\it not} much greater than $a$, then for a very large membrane it is clearly negligibly small compared to the total radius $R_T$, since the latter must diverge as $L_\perp\to\infty$. Thus to complete our proof that $r_0$ is {\it always} negligible compared to $R_T$, we need only consider the case $r_0\gg a$.

It is also clear that in this limit
\bea
R\approx r \,. \label{Rvalue}
\eea 
Both equations (\ref{arclength}) and (\ref{Rvalue}) follow from the fact that, 
for large $r$, the spiral is very nearly locally a circle of radius $r$. Using 
(\ref{arclength}), we have
\bea
ds=r d\theta. \label{ds11}
\eea
Differentiating (\ref{rspiral})  
w.r.t $\theta$ gives us 
\bea 
dr={a \over 2\pi} d\theta. \label{dr} 
\eea 
Combining this along with (\ref{Rvalue}) and
(\ref{arclength}) in (\ref{energysp}), we obtain
\bea
E &\approx& L_{||} \int^{r_{\rm max}}_{r_0} dr \left[{2\pi r \over a}
\left(-{C \over r} + {\kappa_0 \over 2r^2} \right) \right] \nonumber \\
&=& {2\pi L_{||} \over a}\left[-C(r_{max}-r_0) + {\kappa_0 \over 2}\ln({r_{max} 
\over r_0}) \right], \label{energy2}
\eea
where $r_{max}$ is the  radius at which the spiral joins the straight segment connecting it to the other spirals. 
which means that each spiral has a total length of $({L_\bot \over 2}- {r_{\rm max} 
\over 2})$ available to coil up into a spiral.  We can calculate $r_{\rm max}$ by  first taking the ratio of (\ref{ds11}) 
and (\ref{dr}); this gives
\bea
{ds \over dr}={2\pi r \over a}. \label{dsdr}
\eea 
Integrating the above equation over one of the two spirals yields
\bea
{L_\bot - r_{\rm max} \over 2}= \pi{(r_{\rm max}^2-r_0^2) \over a}. \label{lperp}
\eea
Assuming $r_{\rm max}>>a$ (which can be verified a posteriori), 
so that ${r_{\rm max}^2 \over a}>> r_{\rm max}$, we obtain from Eq.~(\ref{lperp}),
\bea
r_{\rm max}=\sqrt{{aL_\bot \over 2\pi} + r_0^2}. \label{rmax}
\eea
Now assuming that $r_0<<\sqrt{aL_\bot}$, (which will also be verified a 
posteriori), Eq.~(\ref{rmax}) can be approximated as
\bea
r_{\rm max}\approx \sqrt{{aL_\bot \over 2\pi}} + \sqrt{{\pi r_0^2 \over 
2aL_\bot}}r_0. \label{rmaxf} 
\eea 
The second term on the right hand side of Eq. (\ref{rmaxf}) is negligible 
compared to $r_0$, provided our assumption of $r_0<<\sqrt{aL_\bot}$ is correct, 
and hence can be neglected, in eq. (\ref{energy2}) for the energy, relative to the 
$r_0$ terms in $(r_{\rm max}-r_0)$. Thus Eq. (\ref{energy2}) can be approximated as
\bea
E={2\pi L_{||} \over a}\left[-C\sqrt{{aL_\bot \over 2\pi}} + C_0r_0 + 
{\kappa_0 \over 2}\ln\left(\sqrt{aL_\bot \over 2\pi r_0^2}\right) \right].\nonumber\\
\label{energy3} 
\eea  
  Minimizing   this expression for the energy over $r_0$ gives us
\bea
r_0={\kappa_0 \over 2C}. \label{r022} 
\eea
We note that since this is independent of $L_\bot$, $r_0$ 
will always be $<<\sqrt{aL_\bot}$ (as assumed above) if $L_\bot>> {\kappa_0^2 \over 
C^2 
a}$. Also since (\ref{rmaxf}) gives $r_{\rm max}\approx \sqrt{{aL_\bot \over 
2\pi}}$, we see that the {\it a posteriori} assumption $r_{\rm max}>>a$ will always be 
satisfied provided $L_\bot>>a$, which it must be for us to describe our system as a membrane. 
Thus all our  assumptions are well-validated. Notice Eq.~(\ref{r022}) 
together with Eq.~(\ref{kappa})
yields that $r_0$ can be tuned to zero by tuning $\chi$. 

We lastly comment on the expansion of $r_0$ due to thermal fluctuations. Since $L_H(s)$ has a 
monotonic dependence on $d(s)\propto s^{1/3}$, for the innermost turns of the spiral near its 
core,  we note that $L_H$ is very small. This means $L_H\ll \xi_{NL}$ for the first few layers, and 
hence the nonlinear fluctuation-corrections to $\kappa_0$ are  small. Thus even at finite $T$, the effective bend modulus for the first few turns of the 
spiral is same as $\kappa_0$, and $r_0$ is still given by (\ref{r022}). This also means the shape 
of the spiral near its core is given by  Eq.~(\ref{rtheta1}) with $\eta=0$ and $\nu=1$. Thus the form of the spiral  near its core in the stable spiral phase with $\chi^2 < \chi_L^2$ is different from 
its shape near the outermost turns.

\bibliography{tethered}

\end{document}